\documentclass[11pt,preprint]{aastex}
\usepackage{epsfig}
\usepackage{color}

\newcommand{\etal}{et~al. }

\newcommand{\parder}[2]{\frac{\partial{#1}}{\partial{#2}}}
\newcommand{\st}{\scriptscriptstyle}
\newcommand{\msun}{{\rm M}_\odot}
\newcommand{\yr}{\,{\rm yr}}
\newcommand{\pyr}{\,\yr^{-1}}
\newcommand{\lsun}{L_{B,\odot}}

\newcommand{\lumin}{{\rm ergs\,s^{-1}}}
\renewcommand{\baselinestretch}{1.4}
\begin{document} 
\title{Type-Ia Supernova-driven Galactic Bulge Wind}
\author{\vspace{-\parskip}
Shikui Tang$^1$, Q. Daniel Wang$^1$,
Mordecai-Mark Mac Low$^{2,3}$,
and M. Ryan Joung$^{4}$}
\affil{\vspace{-\parskip}
$^1$Department of Astronomy, University of Massachusetts,
  Amherst, MA 01003 \\
$^2$Dept. of Astrophysics, American Museum of Natural History \\
$^3$Dept. of Astronomy, Columbia University \\
$^4$Dept. of Astrophysical Sciences, Princeton University
\vspace{-\parskip}
}
\begin{abstract}
\vspace{-8ex}
Stellar feedback in galactic bulges plays an essential role 
in shaping the evolution of galaxies.
To quantify this role and facilitate comparisons with X-ray observations,
we conduct three-dimensional (3D) hydrodynamical simulations
with the adaptive mesh refinement code, FLASH,
to investigate the physical properties of hot gas
inside a galactic bulge, similar to that of our Galaxy or M\,31.
We assume that the dynamical and thermal properties of the hot gas
are dominated by mechanical energy input from supernovae (SNe),
primarily Type Ia, and mass injection from evolved stars
as well as iron enrichment from SNe.
We study the bulge-wide outflow as well as the SN heating
on scales down to $\sim\!4$ pc. An embedding scheme that is
devised to plant individual SN remnant (SNR) seeds,
allows to examine, for the first time,
the effect of sporadic SNe on the density, temperature, and iron ejecta
distribution of the hot gas as well as the resultant X-ray morphology
and spectrum. We find that the SNe produce a bulge wind
with highly filamentary density structures and patchy ejecta.
Compared with a one-dimensional (1D) spherical wind model,
the non-uniformity of simulated gas density, temperature,
and metallicity substantially alters the spectral shape and
increases the diffuse X-ray luminosity.
The differential emission measure as a function of temperature
of the simulated gas exhibits a log-normal distribution,
with a peak value much lower than that of the corresponding 1D model.
The X-ray luminosity depends sensitively on the mass loss
rate from evolved stars. The bulk of the X-ray emission
comes from the relatively low temperature and low abundance
gas shells associated with SN blastwaves. SN ejecta are not well
mixed with the ambient medium, at least in the bulge region.
These results, at least partly, account for the
apparent lack of evidence for iron enrichment in the
soft X-ray-emitting gas in galactic bulges
and intermediate-mass elliptical galaxies.
The bulge wind helps to explain the ``missing'' stellar
feedback in such galaxies.
But the resultant diffuse emission is more than one order of magnitude
less than that observed in the Galactic and M\,31 bulges,
indicating that gas in these bulges is in a subsonic outflow state
probably due to additional mass loading to the hot gas and/or due to
energy input rate that is substantially lower than the current estimate.
\vspace{-\parskip}
\end{abstract} 
 
\section{Introduction} \vspace{-\parskip}
Bulges of early-type spirals
and elliptical galaxies comprise primarily old low-mass stars,
which account for more than half of the total stellar mass in
the local Universe (Fukugita, Hogan, \& Peebles 1998).
These stars collectively generate a long-lasting feedback
via mass injection from evolved stars
(mainly red giant branch stars and planetary nebulae)
in the form of 
stellar winds and energy input from Type Ia SNe.
Because of the SN heating, the ISM should be mostly in
X-ray-emitting plasma inside galactic bulges,
where little cool gas is present (e.g., \citealt{MB1971, Sage07}).
Observations have shown that the X-ray-inferred gas mass and energy
are far less than those empirical predictions
(e.g., \citealt{Sato99, David06}).
In other words, the bulk of stellar feedback expected
is not observed \citep{Wang07}.
This ``missing'' stellar feedback problem becomes particularly acute
in so-called low $L_X/L_B$ (i.e., the ratio of X-ray luminosity to blue
band luminosity) bulge-dominated galaxies
(typically Sa spirals, S0, and low mass ellipticals).
After removing the contribution from point sources
in those relatively deep {\sl Chandra} observations,
the remaining ``diffuse'' X-ray component generally shows a soft spectrum, 
indicating a thermal origin \citep{Irwin02,OEPT03},
and its luminosity is only a few percent of the expected
Type Ia SNe energy input \citep{Li07a, Li07b}. The inferred total mass of
the X-ray-emitting gas falls far short of what is deduced from the
stellar mass loss over the galaxy's lifetime \citep{David06}.

The presence of a bulge-wide outflow may solve the
``missing'' stellar feedback problem (e.g. \citealt{Tang08}).
The 1D solution of an SN-heated bulge outflow, however, has problems.
The physical state of the gas outflow mainly depends on the mass and
energy input rates as well as the gravitational field.
Within those low $L_X/L_B$ galactic bulges of typical mass and
energy input rates, a hot bulge wind\footnote{Hereafter
in our paper we use the term {\sl wind} specifically for
a supersonic outflow and {\sl outflow} for an outflow
in either supersonic or subsonic state.}
should be present theoretically (e.g., \citealt{MB1971,Ciot91}).
However, observations of the diffuse hot ISM are apparently
at odds with the theoretical wind models in nearly all aspects.
Firstly, the predicted X-ray luminosity in a bulge wind scenario is
a few orders of magnitude smaller than the observed (e.g., \citealt{Tang08}).
Secondly, the expected wind temperature is about $\sim$1\,keV or higher,
while the observation-inferred gas temperatures are substantially
lower (e.g., \citealt{David06,Li07a}).
Thirdly, the estimated X-ray surface brightness profile of the wind should be
steeper than that of starlight \citep{Ciot91}, but the observed profiles of
diffuse emission distributions are fairly extended in most low
$L_X/L_B$ bulges or ellipticals
(e.g. fig.~7 in \citealt{Sarazin01}; fig.~5 in \citealt{Li07b}).
Furthermore, the predicted mean iron abundance of the diffuse hot gas
is 3--7 times solar because of the Type Ia SN enrichment
(e.g., \citealt{Ciot91,Sato99}), whereas the observed spectra usually
indicate near- or sub-solar iron abundance.

Some, if not all, of the listed discrepancies may arise from various
oversimplifications of the existing 1D models
(e.g., \citealt{MB1971,WC1983,LM1987,Ciot91}).
In such models the mechanical energy input of SNe is always treated
as pure thermal energy smoothly injected into the ISM.
In reality, however, SNe, sporadic in both time and space,
should naturally produce inhomogeneity in the ISM.
The density and temperature inhomogeneity may significantly affect
the X-ray spectrum and luminosity,
which are proportional to the density square. 
Explosive energy injection in a hot tenuous medium can be
transported away in form of sound wave, so the SN heating
is not local (e.g., \citealt{Tang05}). 
Furthermore, whether or not the SN ejecta of each individual
SNR can be well mixed with the surrounding
material is crucial to address
the apparent low abundance puzzle of the hot gas.
These effects need to be quantified in order to
correctly interpret the existing observations.

In this work, we present a pilot study to explore
the properties of hot gas in galactic bulges by
conducting 3D hydrodynamic simulations.
In these simulations, SNe are randomly generated that statistically
and spatially follow the stellar light distribution.
Based on the mean temperature and density of the surrounding medium,
we adaptively determine the appropriate sizes of individual SNRs
and generate their density, temperature, and velocity profiles
from a library of 1D simulated SNR templates.
We then plant such structured SNR seeds into the 3D simulation grid and
let them evolve. We terminate the simulation when it reaches
a statistically steady state.
The 3D simulations not only provide us with the
dynamical structures of SNRs, but also enable us to trace the
thermal and chemical states of the bulge wind material.

The organization of the paper is as follows.
In \S2 we describe the main physical ingredients of
the bulge wind model and the numerical methods.
The results are presented in \S3 and discussed in \S4.
We summarize our results and conclusions in \S 5.

\section{ Model and Method} \vspace{-\parskip}
\subsection{Model Basics}  \vspace{-\parskip}
We model hot gas inside a galactic bulge that originates
from continuous stellar injection
in the form of stellar winds, and is heated by sporadic Type Ia SNe.
The dynamics of the hot gas is described by the following equations:
\begin{eqnarray}
&\ &\parder{\rho}{t} + 2 \nabla\cdot(\rho\mathbf{v}) = \dot{\rho}_*(r)+
\dot{\rho}_{\st SN}(\mathbf{r},t), \\
&\ &\parder{\rho\mathbf{v}}{t} + \nabla\cdot(\rho\mathbf{v}\mathbf{v})+
\nabla P = -\rho\nabla\Phi, \\
&\ &\parder{\rho E}{t} + \nabla\cdot[(\rho E + P)\mathbf{v}]=
-\rho\mathbf{v}\cdot\nabla\Phi + S_{\st SN}(\mathbf{r},t) + S_{*}(r)
-n_t n_e \Lambda(T), \\
&\ & P  =  n k T
\end{eqnarray}
where $\rho$, $\,\mathbf{v}$, $\,P$, $T$, and $E$ denote density, velocity vector,
pressure, temperature, and total specific energy of the hot gas;
and $\Phi$ is the gravitational potential field;
$n_t$ and $n_e$ are the number density of ions and electrons;
$n=n_t + n_e$ is the total number density; $\Lambda(T)$ is the normalized
cooling function taken from \citet{Sutherland93}, assuming an
optically thin plasma with solar abundance;
$\dot{\rho}_*$ and $S_*$ denote the mass and energy
input from evolved stars.
The mass and energy input from individual SNe,
$\dot{\rho}_{\st SN}(\mathbf{r},t)$ and $S_{\st SN}(\mathbf{r},t)$,
are explicitly expressed as a function of position and time.

We adopt parameters appropriate to the bulge of the Milky Way.
The stellar mass distribution of the bulge follows the
potential-density pair of the Hernquist profile \citep{Hernquist1990}:
\begin{equation}\label{henprof}
\Phi_{\st bulge}(r) = -\frac{GM_{\st bulge}}{r+r_s}, \ \ 
\rho_{\st bulge} (r)=\frac{M_{\st bulge}}{2\pi}\frac{r_s}{r}\frac{1}{(r+r_s)^3},  
\end{equation}
where $r_s$ is the scale radius and $M_{\st bulge}$ is the total mass of the bulge.
Here we set the $M_{\st bulge}$ and $r_s$ to be $2.4\times 10^{10}\,\msun$
and 0.42\,kpc (e.g., \citealt{Kent1992, Zhao1994, Blum1995, Wolfire1995, LepineL00}).
Other components of our Galaxy, such as the disk and dark matter halo,
have only little effects on the gas dynamics within the Galactic bulge
and are thus ignored in our simulations.

The stellar mass loss from evolved stars,
$\dot{\rho}_*(r)$, follows the stellar mass distribution.
The total stellar mass loss rate of the bulge ($\dot{M}=\int\!\dot{\rho}dV$)
is constrained by current theoretical predictions and related observations,
although the true value cannot be
observed directly inside the bulge. It is inferred from our knowledge
of the stellar population and evolution, together with observations
of the mass loss of similar stars in the solar neighborhood.
Estimates of the stellar mass loss rate may vary
by more than a factor of two.
Assuming a single stellar population with a standard
Salpeter initial mass function, \citet{Ciot91} found that
the stellar mass loss rate can be approximated as
\begin{equation} \label{eq:mdot}
  \dot{M}=0.25L_{10}t_{10}^{-1.3}\msun\pyr,
\end{equation}
where $L_{10}$ is the current optical blue-band luminosity
in units of $10^{10}\lsun$ of the stellar population and
$t_{10}$ is its age in units of 10$\,$Gyr.
Adopting a blue-band luminosity of
$2\times 10^{9}\lsun$ (\citealt{Cox00}, p571) and an age of 10\,Gyr
for the bulge, we have $\dot{M} \simeq 0.05\,\msun\pyr$.
\citet[fig.~22]{Mar05} directly relates mass loss to the total mass
of a stellar population (see also \citealt{Tang08}),
which gives $\dot{M} \simeq 0.07 \rm M_\odot yr^{-1}$.
Another estimate based on observations of asymptotic giant
branch stars gives $\dot{M}=0.64 L_{10}$\,$\msun\rm yr^{-1}$,
consistent with the stellar mass loss rate inferred for a sample of
nine ellipticals from mid-IR observations \citep{Athey02}.
Thus, $\dot{M}$ for the Galactic bulge can be as high as
0.13$\,\msun\pyr$.  We therefore run models with different
mass loss rates, as discussed in \S2.3.

The energy feedback of the stellar bulge is dominated by the mechanical
energy from Type Ia SNe. Following the convention, we assume that each SN
releases $10^{51}\rm~ergs$ mechanic energy.
We take the Type Ia SN rate of E-S0 galaxies to be 0.12 SNu
(SNu is defined as one SN per $\rm 10^{10}\,L_{B,\odot}$
per century; \citealt{CappEva1999}; \citealt{Cox00}, p467),
which gives about one SN per 3000 year in our Galactic bulge.
The energy input from the stellar wind, $S_{*}(r)$,
is assumed to be thermalized to their stellar kinematic
temperature $T_* \equiv \mu m_p \sigma^2 /3k \simeq 3\times10^5\,$K,
corresponding to a stellar velocity dispersion around 100 km$\,\rm s^{-1}$
\citep{Eckart1993}.  Overall this energy is almost negligible
compared with that from SNe.

Each SN also ejects an adopted (Chandrasekhar) mass of 1.4\,$\rm M_\odot$.
Though the total amount of SN ejecta is much less than that of the mass loss
from evolved stars, the SN ejecta contribute most of the metals,
especially iron. In order to trace how these iron-rich ejecta
mix with the stellar wind material, we additionally incorporate
a separate advection equation
\begin{equation}\label{eq:ironfrac}
  \parder{\rho \chi_i}{t} + \nabla\cdot(\rho \chi_i \mathbf{v})=0,
\end{equation}
where $\chi_i$ is the mass fraction of the $i$th component,
with the constraint $\sum_i \chi_i = 1$.
In the present simulations we have two components,
the iron mass and the rest of gas mass.
The iron mass from SNe, assumed to be 0.7\,$\rm M_\odot$ per SN
\citep{Nomoto1984}, is part of the SN ejecta.
The iron mass fraction in stellar wind material
is $f_{Fe,\odot}=0.1\%$ (i.e., the nominal solar).
The iron abundance can trace the SN enrichment.
Any zone with iron abundance greater than the solar
value is enriched by SNe. 
Thereafter we refer the iron ejecta as purely from SNe.

The simulations are performed with FLASH \citep{Fryxell00}, 
an Eulerian astrophysical hydrodynamics code
with the adaptive mesh refinement (AMR) capability,
developed by the FLASH Center at the University of Chicago.
FLASH solves the Euler equations and uses the piecewise-parabolic method
to deal with compressible flows with shocks.
We take the advantage of the AMR capability
to accurately include the heating of individual SNRs.

\vspace{-0.5\parskip}
\subsection{One Dimensional  Model} \vspace{-\parskip}
To help set up the 3D simulations, we first simulate a 1D model.
Assuming that the energy and mass inputs are continuous in time and
spherically symmetric in space, we may simplify Eqs.~(1)--(3) into a 1D problem.
For a specific galactic bulge, the {\it crossing time} ---the time required
for the gas to flow from the center to the outer boundary---
is a few million years (assuming a size of a few kpc for the bulge),
significantly shorter than the evolutionary time scale of
the stellar energy and mass input rates.
Therefore, the problem can be regarded
as time-independent. The bulge outflow can reach a steady state
if radiative cooling does not affect the dynamics.

We run the 1D simulations by using a gas-free initial condition
and by continuing to add corresponding energy and mass in each
zone. FLASH handles the mass and energy inputs
with an operator split method. At each time step it first solves
the Eulerian equations without the source terms,
then explicitly updates the solution to account for
the corresponding source terms.
To properly conserve the mass, momentum, and energy,
we implement this procedure in three steps:
first, we update the gas density according to the amount of mass input
in that step; next, we modify the gas velocity to satisfy
momentum conservation; finally, we modify the gas temperature
to conserve the total energy.
We verified that this implementation can exactly produce
the analytical solution of a star cluster wind \citep{Canto00}.

The system eventually evolves to a steady state.
Such a steady outflow solution can be analytically
derived without including cooling \citep{WC1983}.
If the 1D bulge outflow solution has a sonic point,
a subsonic outflow can then develop into a supersonic outflow
(i.e., a bulge wind).  The final state of such a wind does not
depend on the specific initial condition.
Under certain conditions (e.g., due to significant radiative cooling
or low specific energy input; see \S4.2 for more discussion),
the sonic point may not exist and the gas outflow may be sensitive
to the boundary condition as well as the initial condition.
The use of the outflow (sometimes called zero-gradient)
boundary condition would then introduce an artificial force inserted by
the leveled-off pressure that would produce perturbation,
propagating inwards on a time scale comparable to the crossing time.
This situation would also occur if a simulation region were too
small to include the sonic point (if present).
Thus we use the 1D solution to make sure that the sonic point is
included in the 3D simulation domain (a cubic box).

\vspace{-0.5\parskip}
\subsection{Three Dimensional simulations}
\vspace{-\parskip}

Two 3D simulations are performed to examine the
properties of galactic bulge winds.
Their key parameters are listed in 
Table~\ref{T:para} for quick reference.
The major difference between these two simulations is the mass loss rate:
$0.05\,\msun\pyr$ for Model A (the 3D reference model)
and $0.1\,\msun\pyr$ for Model B,
representing the uncertainty in the mass loss rate (\S2.1)
or extra mass loading expected in galactic bulges
(\S4.2; see also Li \& Wang 2009 in preparation).
The highest spatial resolutions are $\sim$ 3.9 and 4.9\,pc
respectively for the two models. The effective single-grid
resolution of the simulations are $1024^3$ and $2048^3$ zones.
The steady wind flow established
in the 1D model is used as the initial flow with
an iron mass fraction of the solar value ($f=0.1\%$).

In the 3D realizations, SNe explode randomly 
according to a Poisson process with the mean overall rate.
Their spatial distribution statistically follows
the stellar density distribution. 
It would be computationally very expensive, if even possible,
to simulate the evolution of each SNR on sub-parsec scales
within the bulge-wide flow.
Instead, we adaptively plant individual structured SNR seeds
into the 3D simulation grid and then let them evolve.
We do not simply adopt the Sedov solution,
which is generally not appropriate for an SNR evolving
in a hot tenuous medium \citep{Tang05},
especially when the dynamics of the SN ejecta are considered.
According to a scaling scheme detailed in
a separate paper \citep{Tang09},
the structure of an SNR can be scaled from a template SNR simulated in 
a different ambient medium setting.
We have constructed a library of template SNRs from 1D simulations,
assuming a selection of ambient gas temperatures and densities.
Each entry of this library consists of the density, temperature,
and velocity profiles at a particular age and a forward shock radius.

We apply the scaling scheme to dynamically generate
the profiles of each SNR seed. Specifically, we select a spherical
region around each SN location within which the density and
pressure are sufficiently smooth, using the L$\rm\ddot{o}$ner's
error (FLASH User's Guide; \citealt{Lohner87}) as the estimator.
We find that at least 500 zones
(i.e., more than 5 points for the radial profiles)
are needed to reasonably well represent a structured SNR seed. 
This in turn requires the minimum embedding radius to be 
at least 20\,pc, given the spatial resolution that is achievable
in our simulations. With the embedding radius determined,
we calculate the mean density and gas-mass-weighted temperature 
of the enclosed gas to find the most suitable template in the library
and to form the required SNR seed \citep{Tang09}. 

The planting of an SNR seed also takes a few steps.
First we refine the affected region to the highest refinement
level. 
Then we normalize the SNR structures to ensure
the conservations of mass, momentum, and energy within that
region (see Appendix A for details).
Finally, the innermost region that
encloses 0.7$\,\msun$ is traced as the pure iron ejecta of the embedded SNR.
This embedding procedure is well parallelized and allows
for linear scaling up to at least 1024 processors.

\begin{deluxetable}{cccccc}
\tabletypesize{\footnotesize}
\tablecaption{Model Parameters}
\tablewidth{0pt}
\tablehead{
Model&$\dot{M}$ &$\dot{E}_{sn}$ &$r_{\rm sonic}$ &$\Delta \rm L $ & L\\
   &  ($\rm\msun\pyr$) & ($10^{40}\lumin$) & (kpc) & (pc) & (kpc)
}
\startdata
A & 0.05 & 1.1 & 1.0 & 3.9 & 4  \\
B & 0.1  & 1.1 & 1.8 & 4.9 & 10
\enddata
\label{T:para}
\end{deluxetable}

To save computing time, one may simulate only one
octant of the bulge by adopting a reflecting
boundary condition at the surfaces across the center.
A test run, however, shows that
a reflecting boundary condition introduces correlated
wave interactions when an SN explodes near the reflecting boundaries.
This effect is not physical and difficult to quantify.
It is most serious near the bulge center where three reflecting
boundaries intersect and the stellar density,
hence the rate of the SNe, is the highest.
Thus we resort to simulating the whole bulge,
which is centered inside the simulation domain.
However, we only simulate one octant at full resolution,
while the highest resolution in the rest of the grid is degraded by
a factor of four, except for regions where SNRs seeds have just
been embedded. These regions are forced to have the full resolution
in all octants, and are held for $10^5$ years
before returning to the default AMR.
We use refinement estimators acting on the density and pressure
to determine whether a block needs to be refined or derefined,
adopting the default criterion suggested (in FLASH User's Guide,
i.e., refining a block if any estimator is greater
than 0.8 and derefining it if all estimators are less than 0.2).
Regions outside the sonic radius
(which is obtained from the 1D model) are allowed to have
a lower refinement level that gradually decreases with radius.
This approach circumvents the reflection boundary problem
at the expense of about 60\% more computing time, which is acceptable.
In addition, it allows
us to examine the resolution effect within a single run.

\begin{figure}[hbpt]
\begin{center}
\plotone{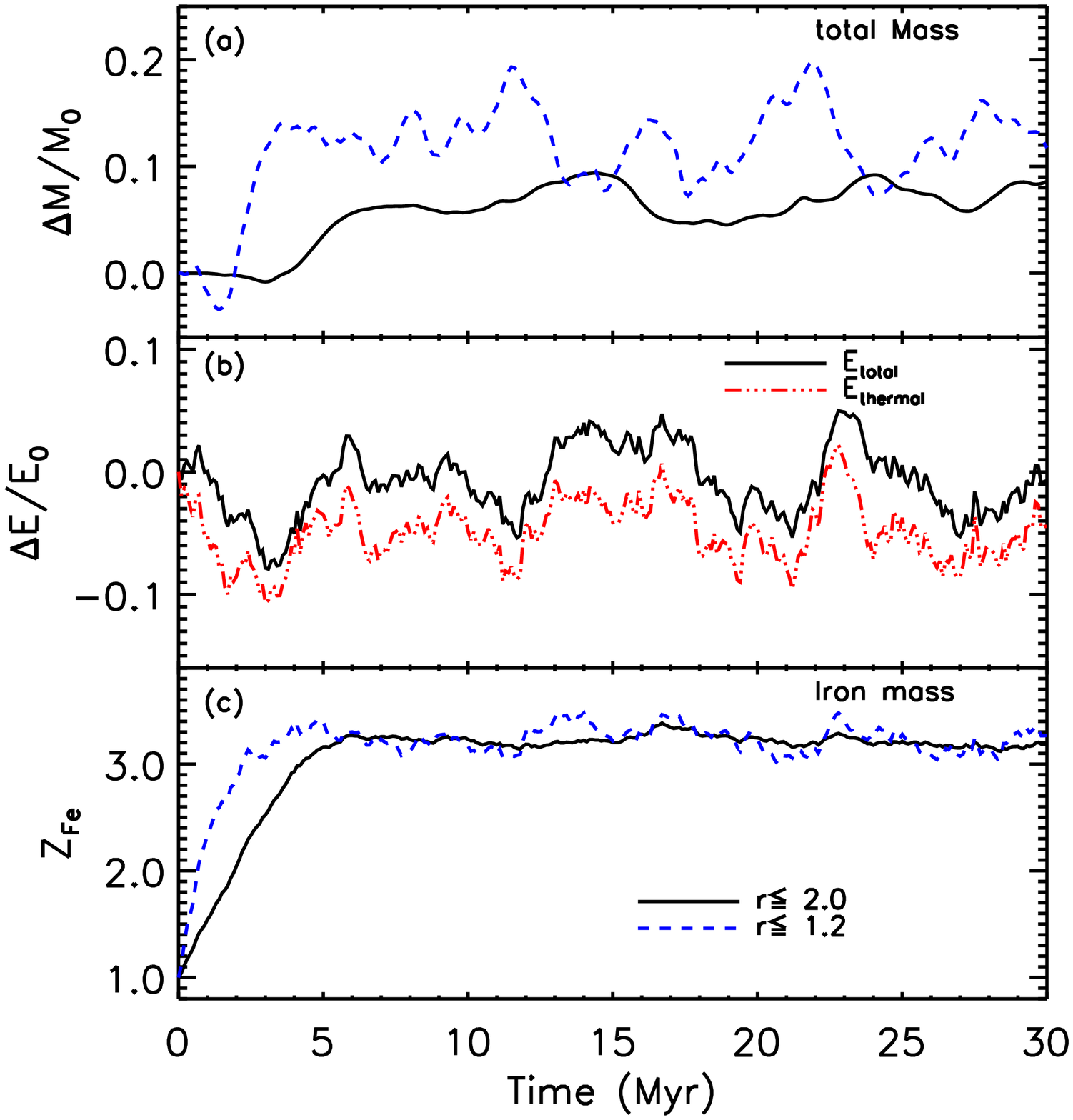}
\caption{\label{F:massen_evol} Variation of the total gas mass [panel (a)]
  and energy [panel (b)] relative to their initial values and iron abundance
  [panel (c)] in Model A. Dash blue lines show the variation within 1.2\,kpc;
  other lines show the variation within 2.0\,kpc.
  The three-dots-dash red line in panel (b) denotes the
  variation of the total thermal energy.
  }
\end{center}
\end{figure}

A statistically steady state of such a 3D simulation can be reached after
a few crossing times.
We quantify the establishment of the steady state of a 3D bulge wind by
examining the relative variation of its global quantities such
as the total mass and energy.
Here we define the variation as the change of
a given quantity relative to its initial value, i.e.,
the relative difference between 3D and 1D.
The variation of the total mass within 2.0\,kpc of Model A is
shown in Fig.~\ref{F:massen_evol}a by the solid line.
Compared to its initial value, the total mass increases to
$\sim 7\%$ on average and fluctuates around this value with a
period of $\sim$\,10\,Myr, comparable to the flow crossing time.
As expected, the mass variation within the inner 1.2\,kpc radius, 
displayed as the dashed line in the figure, has a shorter fluctuation 
period. The expected iron mass fraction,
if fully mixed with stellar wind material, is
0.35\% ($\sim 3.5$ times the solar abundance) in Model A.
We show the variation of the iron abundance (in the solar units)
in Fig.~\ref{F:massen_evol}c.
It takes about 5\,Myr for the hot gas inside 2\,kpc to gain the
expected iron mass. The variation of the iron abundance is smaller than
that of total mass, mainly because it actually only reflects
the ratio of total iron mass to the total gas mass. 
By introducing random SN events in the 3D simulations,
the globally conserved quantities are no longer constant as
they should be in a 1D spherical steady flow.
Only when a hydrodynamic steady state is established in the simulations
(i.e., the fluctuation has reached a statistically stable level) is 
the comparison between 1D and 3D results meaningful.

We check the resolution effect based primarily on X-ray luminosity, which
is particularly sensitive to the density structure in the simulations.
We find that the X-ray luminosity difference between the high resolution octant
and the other seven octants is rather small.
This is partly because the majority of the emission arises from
individual SNRs which are resolved at the same resolution in all the octants.
To examine the resolution effect more directly, we resume Model A
with an increased spatial resolution by a factor of two,
and let it only evolve for 0.1\,Myr, limited by the available computing time.
This simulation produces finer structures of the bulge gas,
and the resultant X-ray luminosity increases about 3\% in the high resolution
octant and about 10\% in the other seven octants.
This demonstrates that our results are quite robust and are only slightly
affected by the spatial resolution.

\section{Results} \vspace{-\parskip}
In this section we present the gas properties extracted from the
3D hydrodynamical simulations. We first detail the results of
Model A (Fig.~\ref{F:bgw0_structure}) and then present Model B
(Fig.~\ref{F:dblm_structure}) for comparison.
Data near the outer region are excluded in our analysis to
avoid any potential artifacts introduced by the assumed outer
boundary condition of the simulations.
We show time-dependent gas properties such as global
structures, individual SNRs, and X-ray luminosities
as well as various time-averaged measurements.
The average is made over a time span ranging from 15 to 30\,Myr,
when the simulations have reached quasi steady states
(see Fig.~\ref{F:massen_evol}).

\subsection{Structures}\vspace{-\parskip}

\begin{figure}[hbpt]
\begin{center}
\plotone{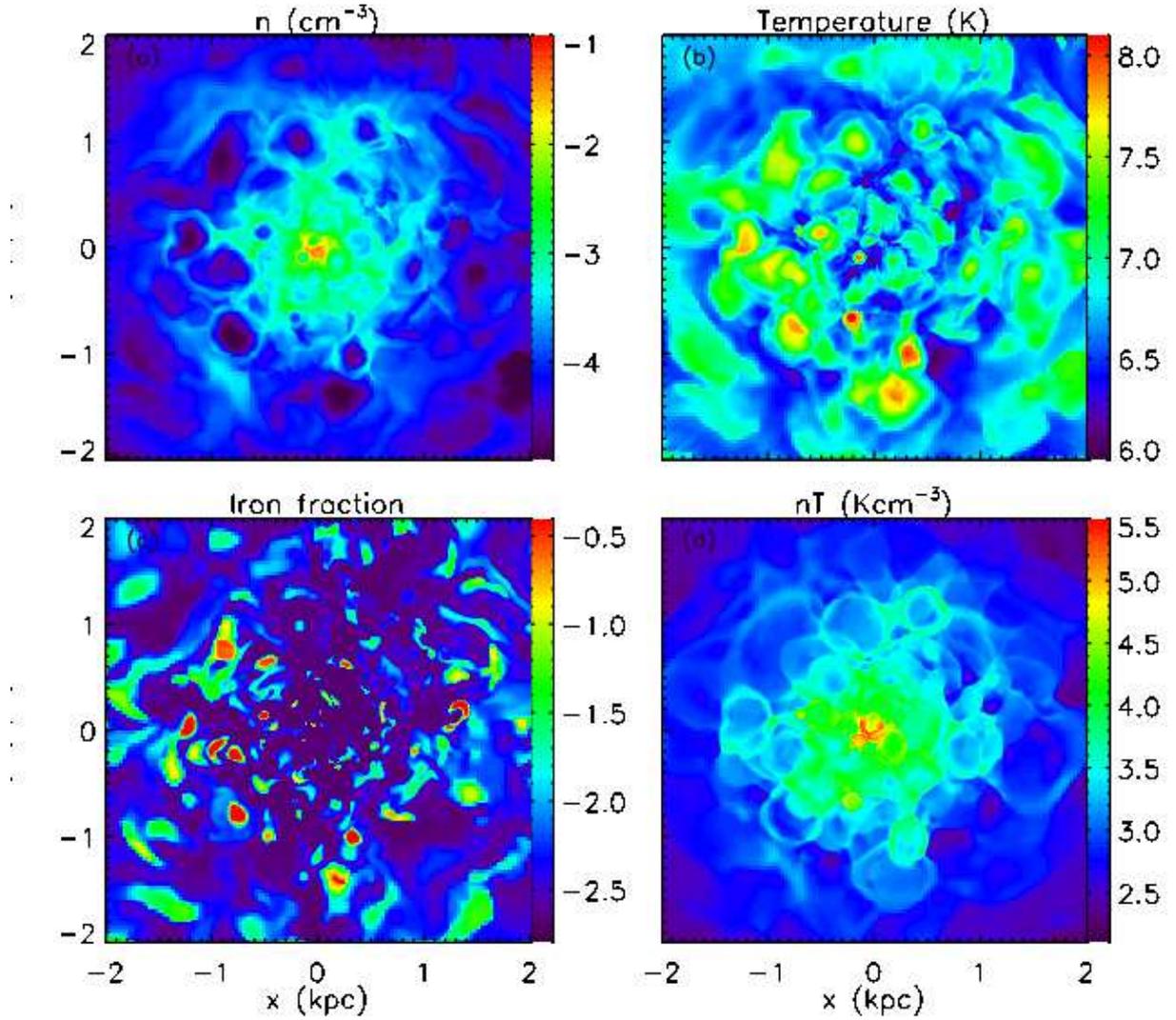}
\caption{\label{F:bgw0_structure}
  Snapshot of Model A in the $z=2$\,pc plane, showing
  (a) density, (b) temperature, (c) iron mass fraction, and (d) pressure.
  All plots are logarithmically scaled according to the color bars.
  Note that the upper right quarter region in each panel denotes
  the data from the octant at full resolution.}
\end{center}
\end{figure}

\begin{figure}[hbpt]
\begin{center}
\plotone{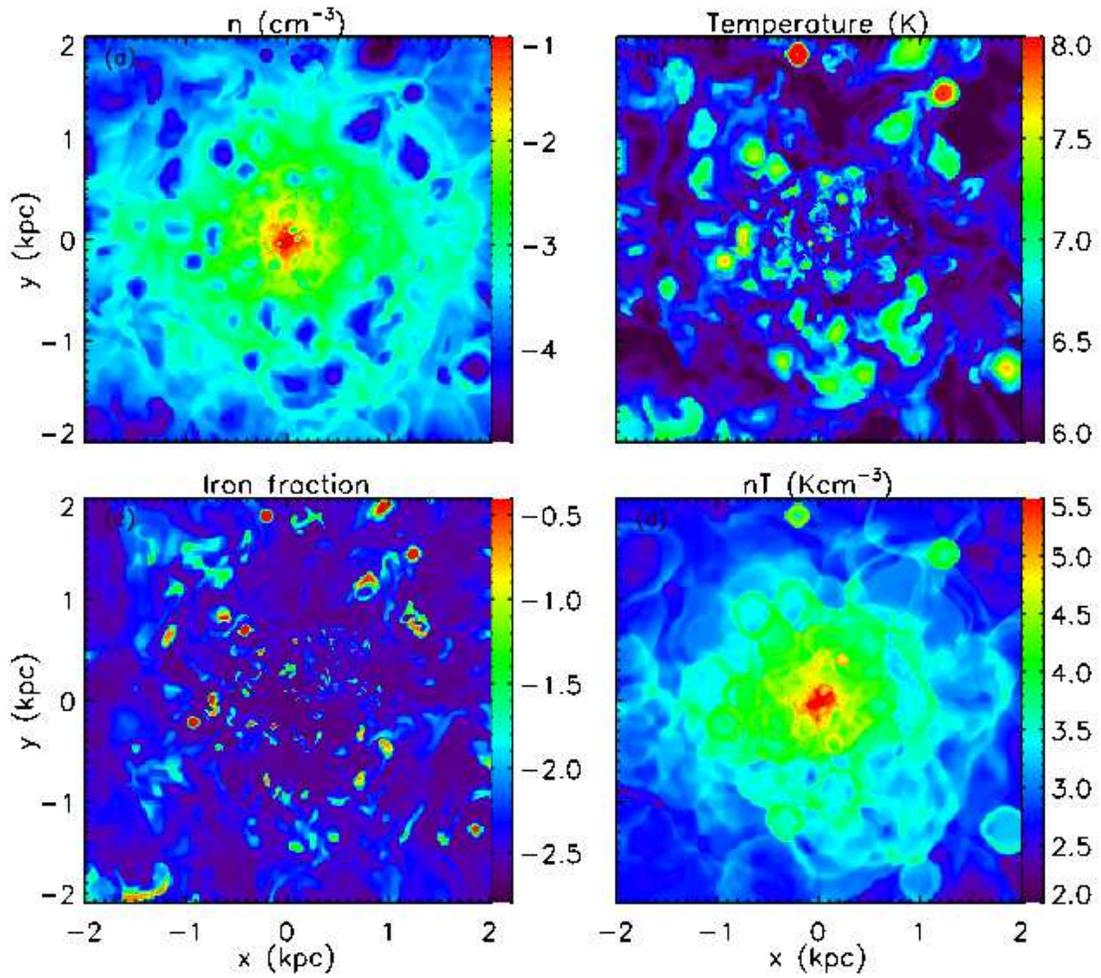}
\caption{\label{F:dblm_structure}
  Snapshot of Model B in the $z=2.5$\,pc plane.
  All plots have the same meaning as in Fig.~\ref{F:bgw0_structure}.
}
\end{center}
\end{figure}

Fig.~\ref{F:bgw0_structure} shows snapshots of the simulated density,
temperature, pressure, and iron ejecta mass fraction of Model A
in the $z=2$\,pc plane.
Sporadic SN events produce non-uniformity in the bulge wind.
The prominent features are various shell-like and filamentary
density structures. These shell-like structures of SNRs are easily
identified in the outer region, where the explosions are less frequent
and each SNR can evolve individually to a large volume
before colliding with others.
Individual SNRs near the bulge center appear more compact
because of the high gas density and pressure
and frequent interactions with adjacent remnants.
Evolved remnants tend to be dispersed and advected outward.
Higher temperature regions in general represent low-density
interiors of SNRs.
The distribution of the pressure is much smoother than those of
the density and temperature (Fig.~\ref{F:bgw0_structure}d),
as has also been found for the SN driven ISM in the Galaxy
\citep{Avillez05,MacLow05,Joung06} and in starbursts \citep{Joung08}.
The shell-like structures in
the pressure map correspond to the expanding blastwaves.
Inside each SNR, the pressure is nearly uniform,
as expected from the \citet{Sedov59} solution.
The spatial distribution of iron ejecta is far from uniform,
as illustrated in Fig.~\ref{F:bgw0_structure}c.
Regions with the lowest iron mass fraction are primarily filled with
stellar wind material with little mixing with the iron ejecta.
Gas with an intermediate iron mass fraction represents iron ejecta
diluted by constantly injected stellar wind material
(or mildly by numerical mixing).
Though being diluted and fragmented, the iron ejecta
is advected out of the bulge, hardly mixing with the bulk of
stellar wind material. Hence, the ISM is not uniformly enriched
by SN ejecta within the bulge.
Similar results are also present in 
Model B (see Fig.~\ref{F:dblm_structure}).

\begin{figure}[hbpt]
\begin{center}
\plotone{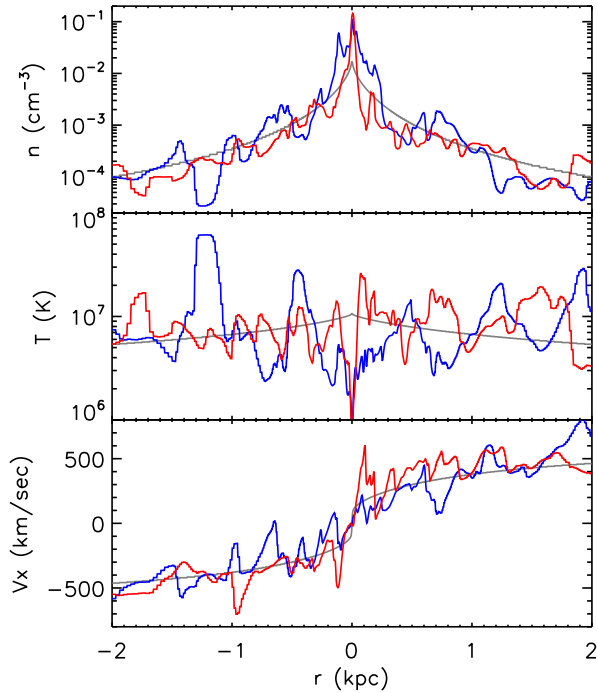}
\caption{\label{F:bgw0_xaxis}
Two sample realizations of density, temperature, and velocity profiles
along $x$-axis. The 1D results (gray lines) are included for comparison.}
\end{center}
\end{figure}

Fig.~\ref{F:bgw0_xaxis} shows sample density, temperature,
and velocity profiles of the hot gas at two representative times.
Individual troughs shown in the density profiles represent interiors
which are surrounded by peaks (i.e., expanding shells of SNRs).
The temperature profiles are nearly
anti-correlated with the density profiles: a peak in temperature usually
corresponds to a trough in density at the same locus,
which is equivalent to the smooth pressure profiles
(e.g., \citealt{MacLow05,Joung06}).
As SNRs evolve, the loci of peaks and troughs change with time.
Multiple evolved SNRs appear to be wave-like.
This wave-like flow driven by sporadic SNe produces the fluctuations
in the globally conserved quantities (Fig.~\ref{F:massen_evol}).

\begin{figure}
\begin{center}
\plotone{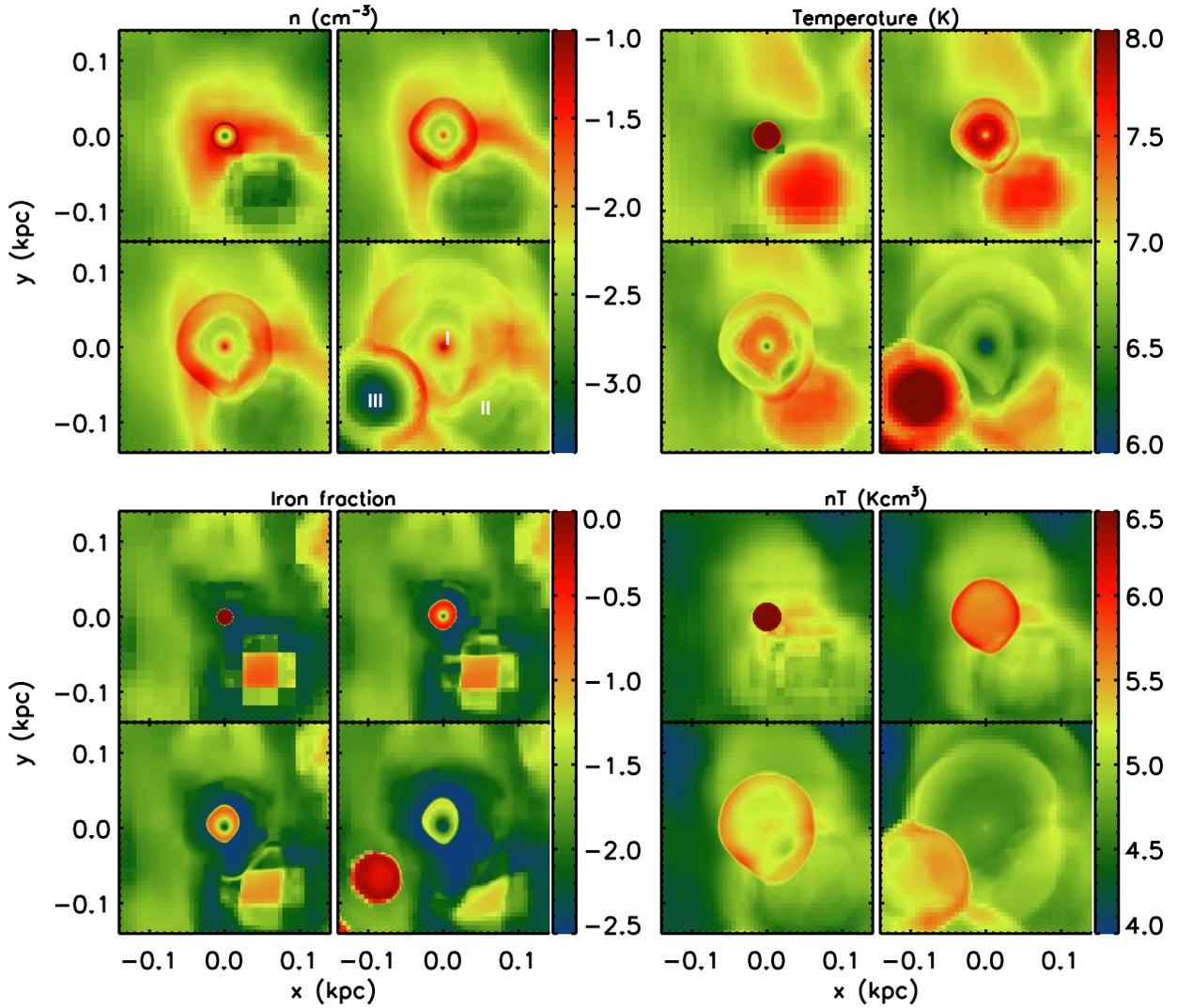}
\caption{\label{F:gcsnr}
  A sample SNR (centered at $\rm I$) evolving near the bulge center.
  The density, temperature, iron fraction, and pressure
   are grouped into four large panels; each panel gives
   snapshots at four ages:
   $10^3$, $2.0\times10^{4}$, $4.0\times 10^{4}$, $9.0\times10^{4}$ year,
   ordered from left to right and top to bottom.
  All values are logarithmically scaled according to the color bars.}
\end{center}
\end{figure}

We demonstrate the evolution of a few SNRs
(labeled as $\rm I$, $\rm I\!I$, and $\rm I\!I\!I$)
very close to the bulge center in Fig.~\ref{F:gcsnr}
taken from the high resolution study.
The forward blastwave of SNR $\rm I$ is evident in the density panel
at $10^3$\,year.
The pure iron core has a radius of about 10\,pc at this time.
The blastwave expands faster toward the lower-right region,
where a lower density cavity has been created by an earlier
SNR (labeled as $\rm I\!I$).
SNR $\rm I$ is unusual in that it occurs right at the bulge center.
Because of the high stellar wind injection rate there, its iron ejecta
are diluted quickly. SNRs away from the center are less affected
by the stellar wind dilution. SNR $\rm I\!I$, for example,
at an age of $2\times 10^4$ year still has an iron mass
fraction of $\sim 6\%$, corresponding to 60 times the solar abundance.
The collision between SNR $\rm I$ and $\rm I\!I\!$
is also evident in the lower-right subpanel of each group.
The evolved shapes of these SNRs are asymmetric
because of both the inhomogeneous environment and
interactions with other SNRs.

\subsection{Time-Averaged Distributions}\vspace{-\parskip}

\begin{figure}[hbpt]
\begin{center}
  \plottwo{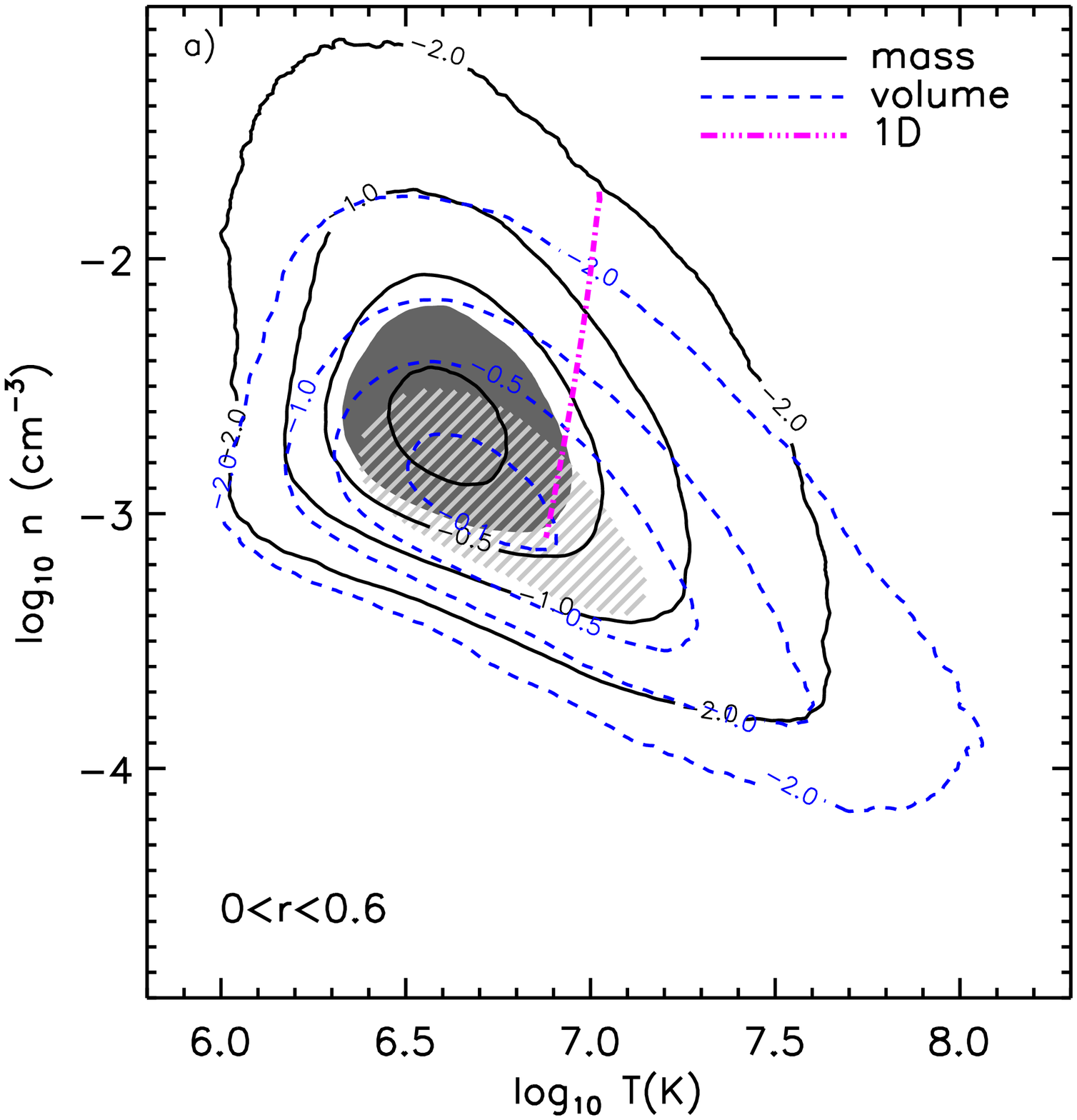}{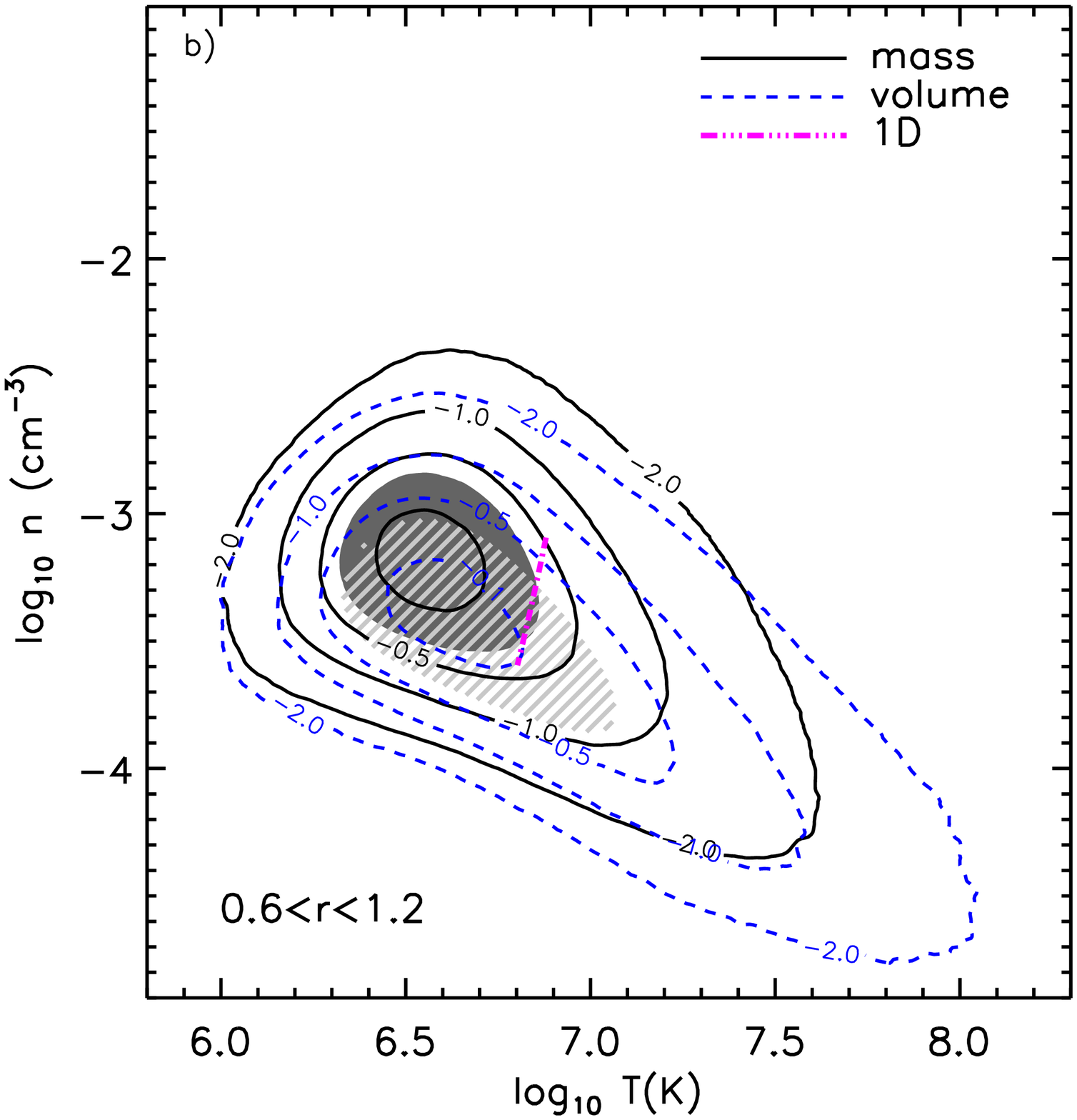}
  \caption{\label{F:bgw0_phase}
    Gas mass (solid black) and volume (dash blue)
    distributions in the temperature-density space in two regions:
    $r\leq 0.6$\,kpc (left panel) and $0.6<r<1.2\, \rm kpc$ (right panel).
    Contours are normalized to their peak values and labelled logarithmically.
    The shaded regions denote the 50\% coverage for
    the mass (solid shading region) and volume (hatched shading region)
    centered on their respectively peak values.
    The corresponding results of the 1D model are shown by
    the three-dots-dashed magenta line in each panel.
  }
\end{center}
\end{figure}

\begin{figure}[hbpt]
\begin{center}
  \epsfig{file=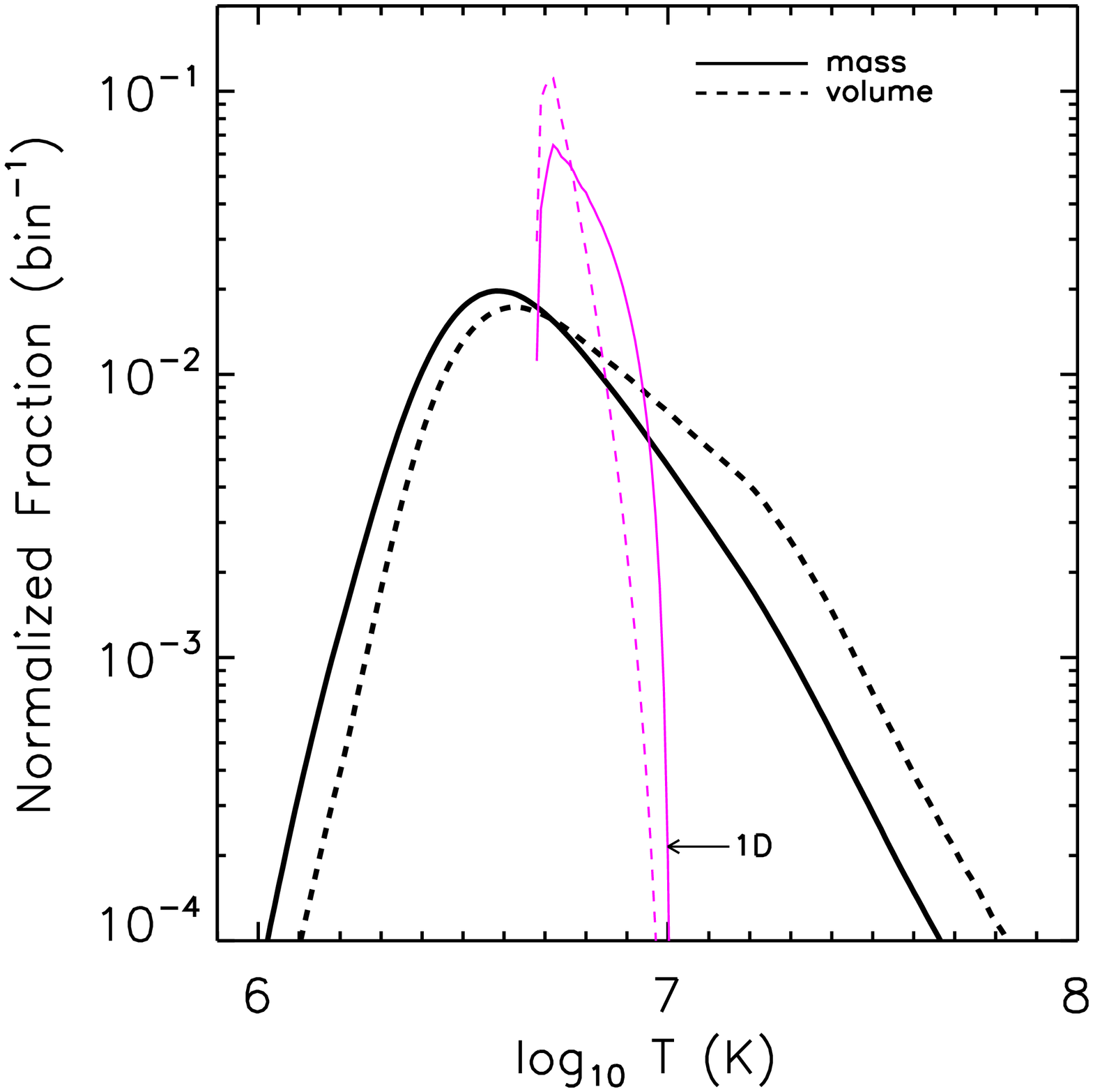,width=0.45\textwidth}\epsfig{file=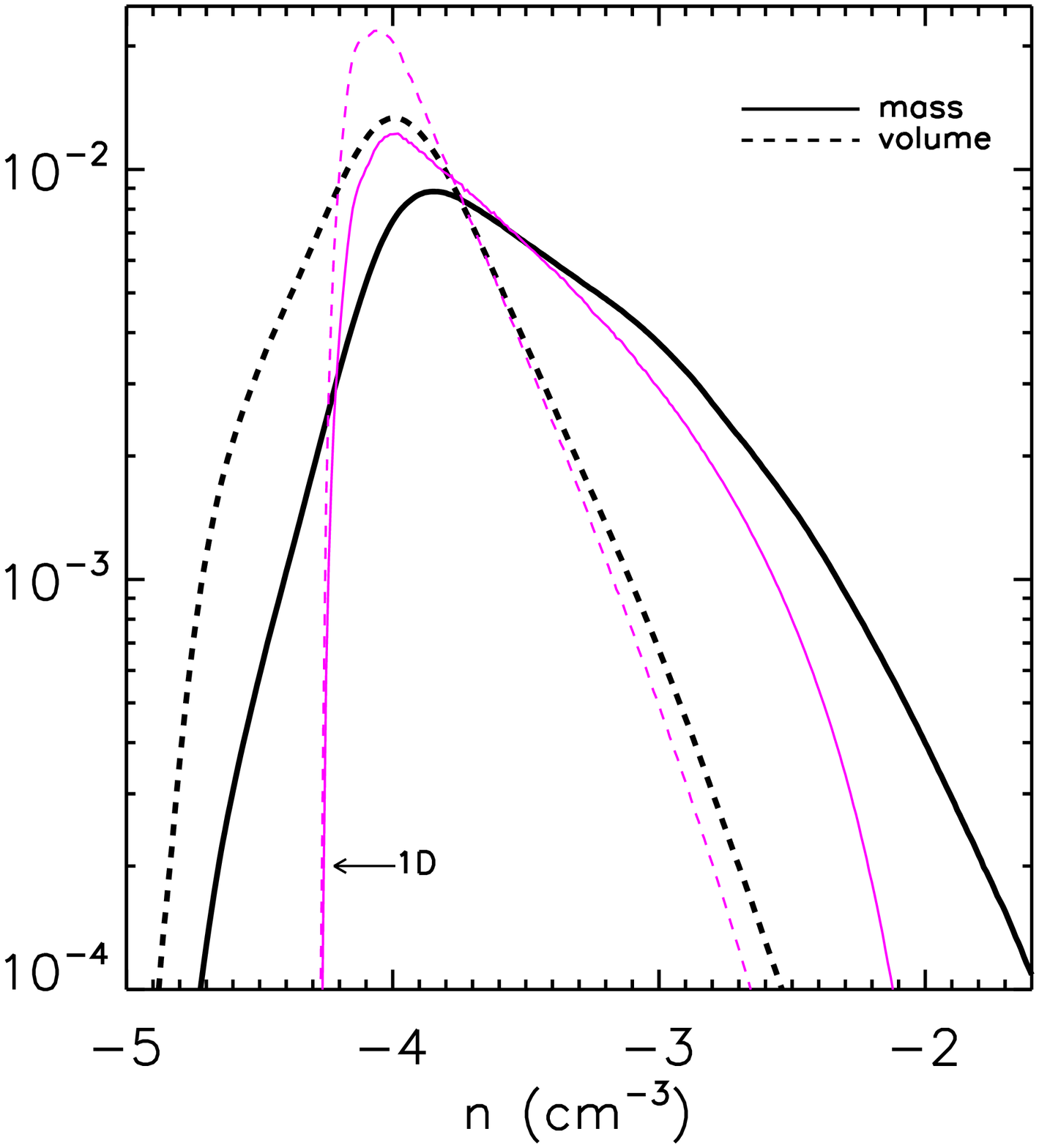,
    width=0.45\textwidth}
  \caption{\label{F:bgw0_pdfnt}
    Gas mass (thick solid) and volume (thick dash)
    distributions as functions of temperature (left-panel) and
    density (right-panel) within $r<1.2\,\rm kpc$.
    The corresponding distributions of the 1D model are shown in
    thin magenta lines. 
 }
\end{center}
\end{figure}

Fig.~\ref{F:bgw0_phase} shows the time-averaged gas mass and volume
distributions in the high resolution octant
as functions of temperature and density in two
regions: $0<r<0.6$\,kpc and $0.6<r<1.2$\,kpc.
The distributions in both regions are broad.
As expected, the mass distribution is biased toward
the lower temperature and higher density side
relative to the volume distribution. Relatively cold dense gas
(e.g., distributed in the upper-left corner of Fig.~\ref{F:bgw0_phase}a)
occupies negligible volume, lying outside
the 1\% contour level of the gas volume distribution.
To see the results more quantitatively, we show their marginal
distributions in temperature and density respectively
(Fig.~\ref{F:bgw0_pdfnt}).
The volume distribution resembles a power law
with an index of roughly $-$2 at the high-temperature end.
The mass distribution is similar to the volume
distribution but peaks at a lower temperature.

\begin{figure}[hbpt]
\begin{center}
  \plotone{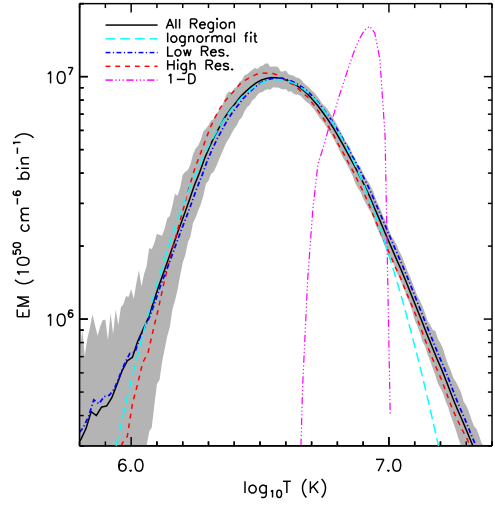}
  \caption{\label{F:bgw0_dem} 
    Time-averaged EMs as a function of temperature:
    the whole simulation region (solid black line);
    the low resolution octants (dash-dot blue line);
    the high resolution octant (short dash red line);
    and the 1D model (three-dots-dashed magenta line).
    The EMs of the low and high resolution regions are linearly
    scaled to the whole region according to their respective volumes.
    The grey region marks the 50\% range around
    the mean EM of the whole region.
    Long-dash cyan line is a log-normal fitting with the mean 
    $\log_{10} T = 6.57$ and the standard deviation of 0.24.
    The bin size of the temperature is 0.01 on the logarithmic scale.
 }
\end{center}
\end{figure}

Fig.~\ref{F:bgw0_dem} shows the differential emission measure (EM)
as a function of temperature within a radius of 1.2\,kpc.
The EMs from the low- and high-resolution
regions are nearly identical around the peak value.
The bulk of the broad EM distribution,
peaked at $\sim 3.7\times10^{6}$\,K,
can be approximated by a log-normal distribution,
with small deviations mostly at the high temperature end.
The grey region in the figure encloses the 50\% intervals
below and above the mean of the whole region.

\subsection{X-ray Emission and Spectra\label{emspec}}
\vspace{-\parskip}

\newcommand{\capAa}{1D: $\rm Z_{\odot} / 3.5\,Z_{\odot}$ }
\newcommand{\capAb}{3D: $\rm Z_{\odot} / 3.5\,Z_{\odot}$ }
\newcommand{\capBa}{1D: $\rm Z_{\odot} / 1.8\,Z_{\odot}$ }
\newcommand{\capBb}{3D: $\rm Z_{\odot} / 1.8\,Z_{\odot}$ }
\begin{deluxetable}{c|cccc} 
\tabletypesize{\footnotesize}
\tablewidth{0pt} \tablecolumns{5}
\tablecaption{\label{T:luminosity}Time-averaged X-ray Luminosities}
\tablehead{Enery band(keV) & 0.2-0.5  & 0.5-2.0 & 2.0-10 & Iron K$\alpha$\\
 ($\lumin$) &($10^{36})$ &($10^{36})$ &($10^{36})$ &($10^{32}$)}
\startdata
(A) \capAa & 0.092/0.22 & 0.72/2.42  & 0.022/0.066 & 0.095/0.31 \\ \hline
(A) \capAb & 0.39/1.18  & 1.22/4.14  & 0.021/0.062 & 4.1/13.8\\
 \hline \hline
(B) \capBa & 3.44/5.32 & 10.8/18.5& 0.03/0.05  & \nodata/\nodata \\ \hline
(B) \capBb & 26.8/49.8& 29.09/50.4& 0.066/0.10 & 7.9/13.5\\
\enddata
\tablecomments{For each model the X-ray emissions are calculated
  using two metallicities: 1.0 and 3.5 solar for Model A;
  1.0 and 1.8 solar for Model B.}
\end{deluxetable}

\begin{figure}[hbpt]
  \begin{center}
    \plotone{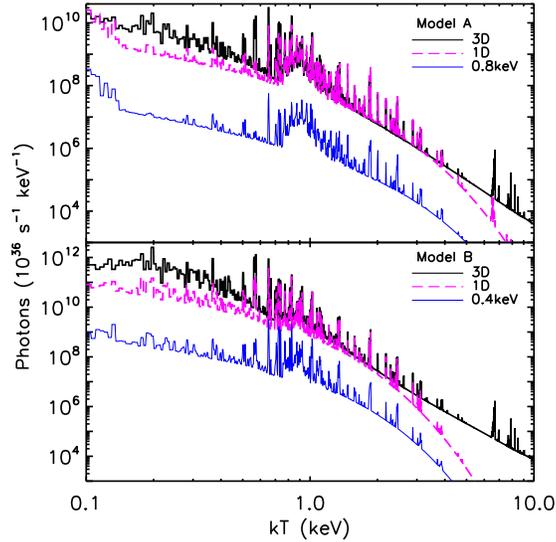} 
    \caption{\label{F:spectrum}
      Time-averaged spectra within $r<1.2$\,kpc
      (solid black line) of Model A (top panel)
      and Model B (bottom panel), compared with
      their corresponding 1D model (dashed magenta line)
      and arbitrarily normalized spectra of isothermal plasma with a
      temperature of 0.8\,keV or 0.4\,keV.
      The solar abundance is used to generate all the spectra.}
  \end{center}
\end{figure}

We calculate the luminosities and spectra using the standard
software package {\small XSPEC} based on the EM distributions.
For the hot gas considered here, we adopt the standard {\small MEKAL} model
\citep{Mewe85,Liedahl95}
for plasma in collisional ionization equilibrium.
Fig.~\ref{F:spectrum} shows a synthesized spectrum of Model A
assuming a solar abundance.
The luminosities in a few bands of Model A
are listed in the 4th row of Table \ref{T:luminosity}.
The corresponding spectra and luminosities of Model B are presented
in Fig.~\ref{F:spectrum} and in Table \ref{T:luminosity} as well.
The luminosity of Model B in the low energy band increases
dramatically; e.g, in the 0.2-0.5\,keV band it
is more than 60 times larger than that of Model A.
This increase is due to the combination of higher density and
lower temperature (hence higher emissivity)
of the weak bulge wind (see \S4.2 discussion).
Of course, the X-ray emission depends on the metal abundance as well.
If an abundance expected for stellar material fully mixed with SN
ejecta were adopted, the luminosities in the three bands would
increase by a factor of a few, as listed in Table \ref{T:luminosity}.

\begin{figure}[hbpt] 
  \begin{center}
    \plotone{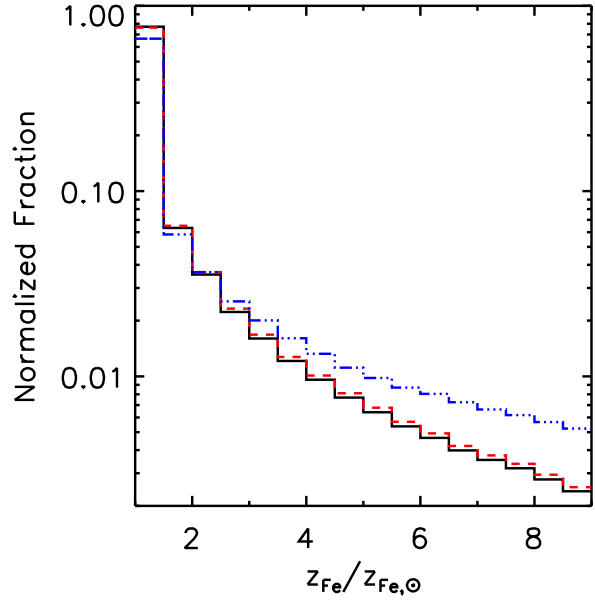}
    \caption{\label{F:emz} Distribution of the emission measure (solid line),
      as well as the luminosities in the 0.3-2.0 keV (dash red line) and
      the 2.0-5.0 keV (three-dots-dash blue line) bands
      as functions of the iron mass fraction.}
  \end{center}
\end{figure}

Since the iron ejecta are in fact not well mixed with
the surrounding material, we examine the effect of
the non-uniform metal distribution on X-ray emission.
The normalized EM distribution as a function of the iron abundance
is plotted in Fig.~\ref{F:emz} as the solid black line.
It shows that about 80\% of the EM comes from the material
with the iron abundance less than 1.5 times solar.
The corresponding distribution of the luminosity in the
0.3-2 keV band nearly follows that of the EM.
But the distribution of the luminosity in the 2.0-5.0 keV band is
affected considerably by gas with higher iron mass fractions
(corresponding to SNR interiors that in general have higher temperatures).
Thus the majority of the X-ray emission from the bulge wind
comes from the stellar wind material
that is hardly enriched by SNe ejecta.
In the following, we thus adopt the solar abundance for
our calculations. 

\begin{figure}[hbpt]
\begin{center}
  \plotone{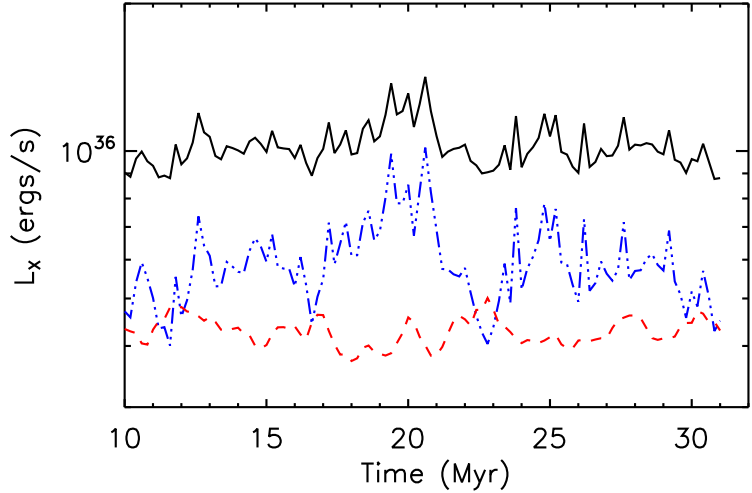}
  \caption{\label{F:bgw0_lxt}
    Illustration of the 0.3-2.0 keV luminosity fluctuation
    as a function of time
    in the region within $r\leq 0.6$\,kpc (three-dots-dash blue line),
    between $0.6\le r \leq 1.2$\,kpc (dash red line),
    and the total volume (solid black line).
 }
\end{center}
\end{figure}

Fig.~\ref{F:bgw0_lxt} shows the time variation of the 0.3-2.0 keV
luminosity of Model A in two concentric regions:
an inner region with $r < 0.6$\,kpc,
and an outer region with $0.6 < r < 1.2$\,kpc.
The luminosity in the inner region has a larger fluctuation
(up to a factor of 3) than that in the outer region.
The X-ray emission from the region with $r>1.2$\,kpc is negligible.
The total 0.3-2.0 keV X-ray luminosity is only
$\sim 10^{36} \rm~ergs\,s^{-1}$ with a fluctuation less than
a factor of two. Thus overall the X-ray emission of the
diffuse gas is not significantly affected by the sporadic SN heating.

\begin{figure}[htbp]
\begin{center}
  \plotone{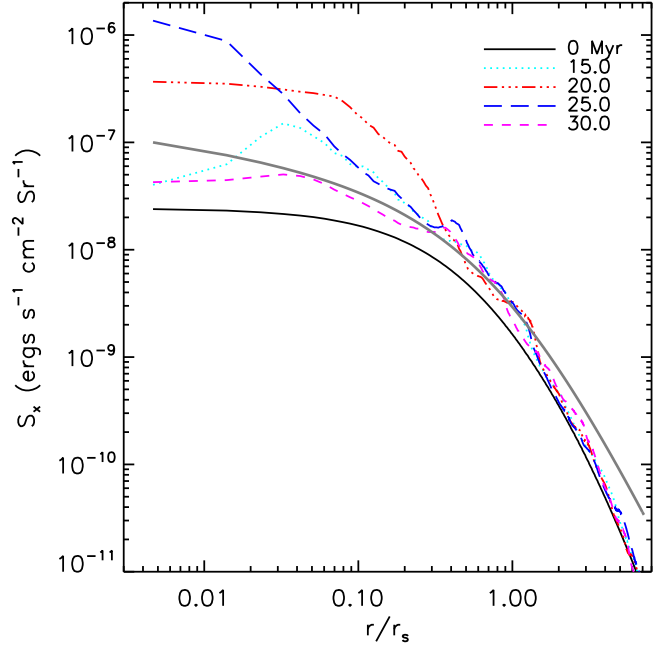}
\caption{\label{F:surbx}
  Sample projected radial intensity profiles in the 0.3-2.0 keV band.
  The gray line shows the assumed stellar surface density distribution
  with an arbitrary normalization.}
\end{center}
\end{figure}

\begin{figure}[htbp]
\begin{center}
\plotone{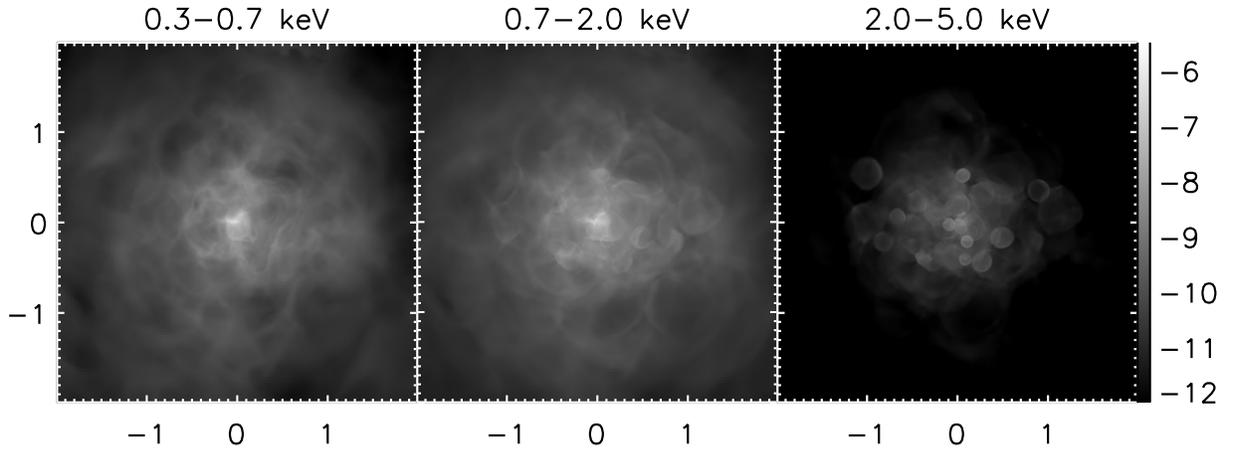}
\caption{\label{F:emap3b}
  Representative intensity maps at full numerical resolution
  in the three energy bands. The unit of $x$ and $y$ axis is kpc.
  The intensity is in the units of $\rm ergs\,s^{-1}\,sr^{-1}$ and
  is logarithmically scaled.
}
\end{center}
\end{figure}

Fig.~\ref{F:surbx} illustrates the 0.3-2.0 keV surface brightness
profile at several representative times.
The brightness varies by more than an order of magnitude near the center.
Compared with the stellar surface density profile (gray line), 
the X-ray profiles are generally steeper.
Fig.~\ref{F:emap3b} shows the X-ray surface brightness maps in
three representative bands (0.3-0.7, 0.7-2.0, and 2.0-5.0 keV).
The maps in the two lower energy bands do not show
significant structures; those small features are typically not
associated with individual SNRs. In the 2.0-5.0 keV map,
individual SNRs are recognizable because they are the primary
source of the hard X-ray emission.

\subsection{Energy Distribution } 
\begin{figure}[hbpt]
  \begin{center}
    \plotone{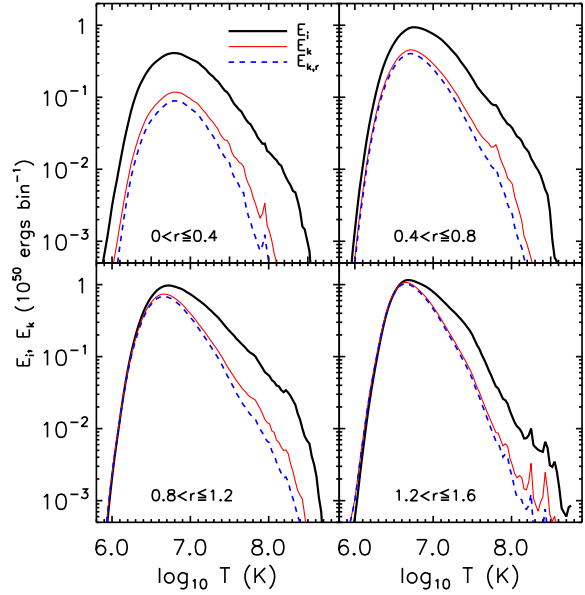}
    \caption{\label{F:energyform}Time-averaged thermal ($E_i$)
      and kinetic energy ($E_k$; $E_{k,r}$ is the radial component)
      distributions versus the gas temperature in four radial ranges
      as marked at the bottom of each panel.}
  \end{center}
\end{figure}

\vspace{-\parskip}
From the 3-D realization, we can further
quantify the thermal, kinetic energies as well as
the turbulent motion of the hot gas.
Fig.~\ref{F:energyform} shows the distributions of
kinetic and thermal energies as a function of temperature
at four different regions.
The energy distribution peaks at $5\times 10^6$\,K.
At temperature below $5\times10^6$\,K,
the energy is dominated by the thermal energy
near the center ($0<r\leq 0.4$\,kpc),
while the thermal and kinetic energies contribute almost equally
at large radii ($r>1.2$\,kpc).
Note that for ideal gas with the ratio of specific heat $\gamma=5/3$,
the Mach number is 1.34 (i.e., supersonic flow)
when its thermal energy equals to
its kinetic energy. The thermal energy of much hotter gas,
primarily low-density hot SNR ejecta,
is always much larger than its kinetic energy, which means that
the SNR ejecta leave the simulation region subsonically.
The difference between the total kinetic energy and its radial
component reveals the energy contained in non-radial motion.
Inside the scale radius ($0<r\leq 0.4$), $\sim$ 30\%
of the kinetic energy is in non-radial motion. This fraction
is less than 2\% in the outer shell ($1.2<r\leq 1.6$).
As a whole, $>$ 80\% of the thermal and kinetic energy of the bulge wind
is stored in gas with temperature between $10^{6.5}$\,K to $10^{7.5}$\,K.
The hotter gas ($T>10^{7.5}$\,K) contains $<$ 5\% of the total energy.

\section{Discussions} \vspace{-\parskip}
\subsection{Comparison between the 1D Model and 3D Simulation}
\vspace{-\parskip}

\begin{figure}[hbpt]
\begin{center}
\plotone{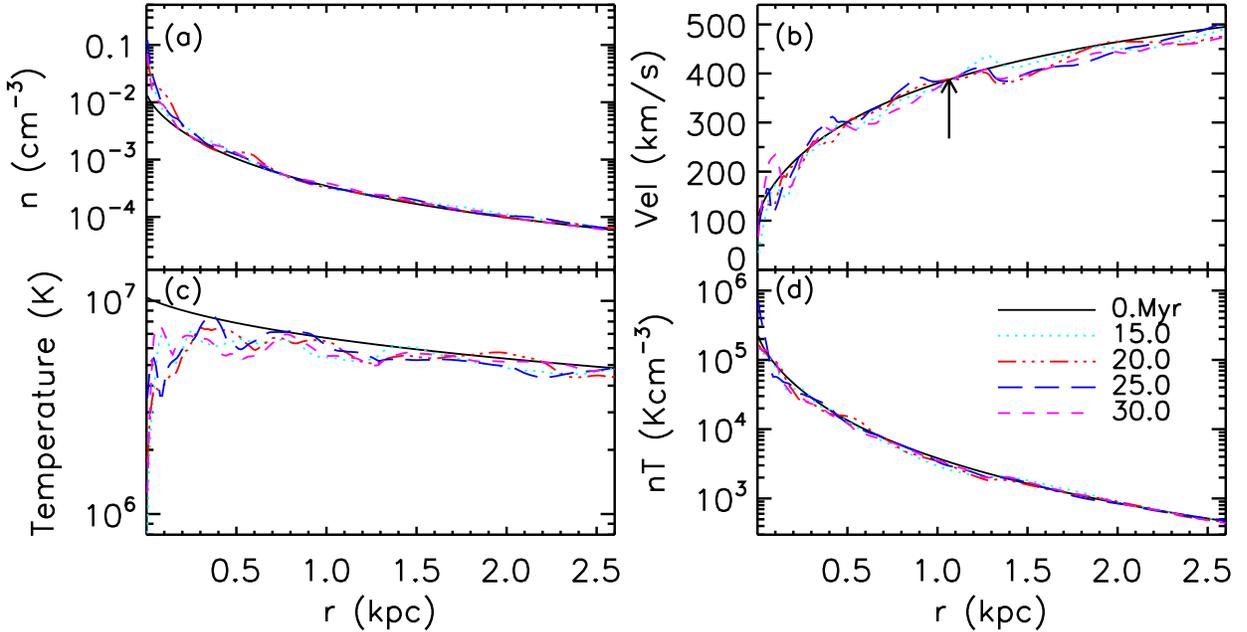}
\caption{\label{F:radial_prof} Radial profiles of density (a),
  velocity (b), temperature (c), and pressure (d) sampled at four times.
  The solid lines ($t$=0.0\,Myr) show the 1D results.
  The density profile is weighted by volume and the other profiles
  are weighted by mass. The arrow in panel (b) marks the sonic point
  of the 1D model.}
\end{center}
\end{figure}

\begin{figure}[hbpt]
\begin{center}
\plotone{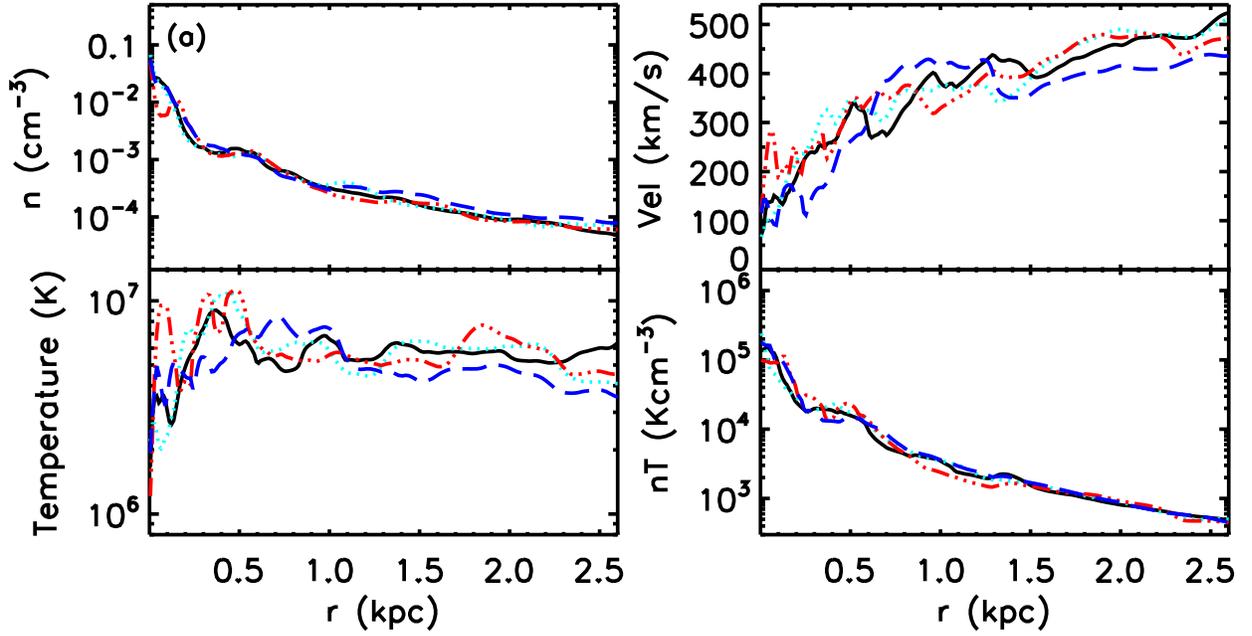}
\caption{\label{F:radial_prof_res} Radial profiles of density (a),
  velocity (b), temperature (c), and pressure (d)
  sampled at four octants.
  The solid black lines show the results of the octant at full resolution,
  and other lines of three octants of low resolution.
}
\end{center}
\end{figure}

Fig.~\ref{F:radial_prof} compares averaged radial profiles for the
density, temperature, velocity, and pressure in a few snapshots of
Model A with the corresponding 1D results.
The velocity and pressure profiles averaged in radial bins follow the
1D results excellently. In the 3D simulation there is no unique
location of the sonic point to divide the supersonic flow from
subsonic flow since the velocities and temperatures vary greatly
from point to point. Nevertheless, the average hot gas outflow
does become supersonic at $\rm \sim 1.0\,kpc$, close to the 1D
sonic point (marked by the arrow).
The density and temperature profiles
deviate significantly from the corresponding
1D results in the inner region ($r<r_s$).
In particular, the density profile tends to be more
centrally peaked in the 3D simulations.
This temporary accumulation of stellar wind material is
largely due to the continuous mass
injection and the lack of prompt heating.
The variation of the density profiles, especially near the center,
is closely coupled with the realization of SN events.
At large radii the density and temperature profiles
become nearly identical to the 1D results.
Since the radiation of the bulge wind is negligible,
the identical radial profiles at large radii
are expected due to the conservation of mass and energy.
Thus the nature of sporadic SN explosions does not affect the
overall gas dynamics on large scales.
In Fig.~\ref{F:radial_prof_res} we compare the
profiles of the octant at full resolution to profiles of three
low resolution octants. The figure suggests that the profiles show
numerical convergence at a level better than the random variance
among realizations.

On average the gas temperature in the 3D simulation has
a lower and flatter profile than that in the corresponding
1D result. In the 1D model, the gas temperature is the highest
at the center and decreases outwards monotonically. In contrast,
the average gas temperature in a 3D simulation can be much lower
at the center than in the surroundings,
because of non-uniformly distributed SN heating.
The distribution of SN heating depends on the ambient medium.
An SN that explodes in a dense environment like the bulge center,
heats a small region to a rather high temperature.
The small amount of overheated gas advects outward
and carries a large fraction of the SN energy with it,
while leaving much of the central gas unheated.
The unheated gas accumulates around the bulge center,
resulting in a relatively steep density profile near the
bulge center (see Fig.~\ref{F:radial_prof} panel a).
The large density gradient makes it even harder for the SN
heating to be uniformly distributed near the center,
so it tends to be transported to outer low density regions
(see also \citealt{Hnatyk99}). Thus a low-temperature inner region
naturally forms under the sporadic SN heating scenario.

\begin{figure}[hbpt]
\begin{center}
\plotone{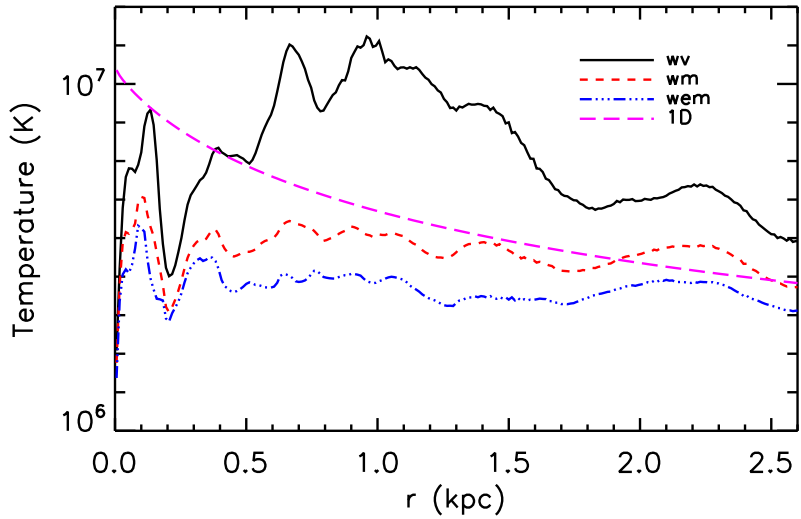}
\caption{\label{F:cmpT}
  A sample radial temperature distribution
  weighted by volume (denoted as wv), by mass (wm),
  and by emission measure (wem).
  The long dash line denotes the 1D temperature profile.
}
\end{center}
\end{figure}

It is also worth noting that the temperature profile depends on
the weighting methods, as shown in Fig.~\ref{F:cmpT}.
The EM-weighted value, which closely resembles that
inferred from X-ray observations, is relatively low because
it is primarily determined by the low-temperature
denser material (see \S3.2).
In comparison, the temperature profiles from different
weighting methods are identical in the 1D model.
This is because in the 1D model, all quantities such as
temperature, density, and pressure are monotonic functions of radius.
Therefore the density has a one-to-one correspondence to the temperature,
which also explains the single-line gas distribution of the 1D model
in the temperature and density space (Fig.~\ref{F:bgw0_phase}).


The integrated gas properties of Model A are also significantly
different from those of the 1D model, such as the EM distributions,
spectra, X-ray luminosities, etc.
In the 1D model the EM peaks at $\sim8\times10^{6}$\,K
and is truncated sharply at low or high temperatures
(three-dots-dashed magenta line in Fig.~\ref{F:bgw0_dem}),
while the EM distribution of Model A peaks at a lower
temperature and is much broader.
But the difference between the spectral shape of Model A and
the corresponding 1D model is small in the 0.7-3.0\,keV band
(see Fig.~\ref{F:spectrum}). Model A has slightly
higher (about 30\% more) X-ray luminosity in this band.
At lower ($<0.7$\,keV) and higher energy bands
($>3.0$\,keV), Model A gives much higher photon fluxes,
especially for some line features, such as OVII (22.0\AA, 0.56\,keV)
and helium-like iron K$\alpha$ ($\rm FeXXV\,K\alpha$; 6.7\,keV),
largely due to the much broader temperature distribution.
It is worth noting that the $\rm FeXXV\,K\alpha$ line emission
in Model B increases but the line in its corresponding 1D model
is missing because of the relatively low and narrow
gas temperature distribution in the 1D model.

\subsection{Observational Implications} \vspace{-\parskip}

There are two direct predictions from the 3D simulations
that help to understand partly the observational puzzles
faced by the 1D wind model.
First, the relatively low emission-weighted gas temperature
is consistent with the results inferred from the diffuse
X-ray emission in galactic bulges and elliptical
galaxies (e.g, \citealt{Sarazin01,David06,Li07a}).
Second, the SN Ia ejecta are not fully mixed with the
stellar wind material inside the galactic bulge, and most of the X-ray
emission is contributed by the stellar wind material shocked by
SN blastwaves. Thus the X-ray spectra in general reflect only
the metal abundance of the stellar wind material,
which is consistent with the apparent solar metallicity inferred
from the diffuse X-ray emission in a large sample of ellipticals
(e.g., \citealt{Humphrey06}). 

\begin{figure}[htbp]
\begin{center}
\plotone{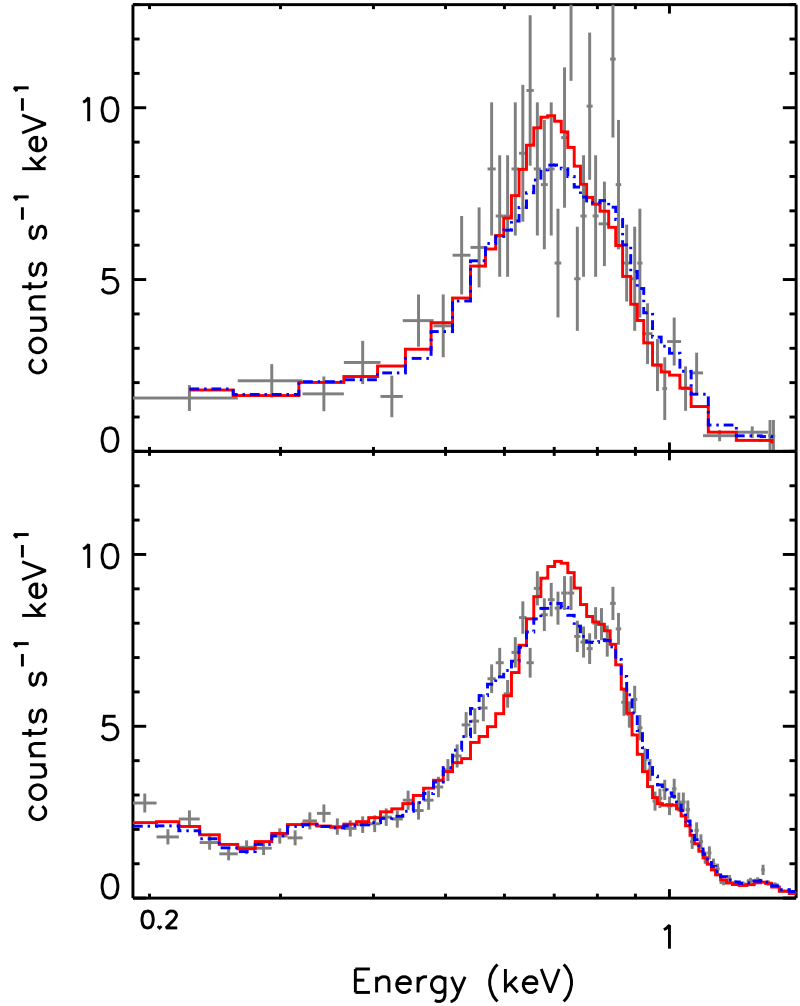}
\caption{\label{F:simspec}
  Simulated {\sl Chandra} ACIS spectra of the X-ray emission
  from the modeled bulge gas, based on the differential EM
  shown in Fig.~\ref{F:bgw0_dem}.
  About 600 and $10^{4}$ photons are generated for
  the upper and lower panels, respectively. The solid red line
  denotes the fit of an isothermal plasma model while the
  dash-dot blue line denotes the simulated spectrum.
  The normalization is arbitrary.
}
\end{center}
\end{figure}

In addition, the broadening of the gas temperature distribution
can also affect the determination of metallicity.
A low metal abundance could be obtained by fitting the X-ray
spectra of such gas with an isothermal plasma model.
This effect is demonstrated in Fig.~\ref{F:simspec}.
We simulate the X-ray spectrum based on the {\small MEKAL} model
in {\small XSPEC}, adopting the approximate log-normal
EM distribution (see \S3.2) and a solar abundance.
The data points in the upper panel
show a simulated spectrum of about $600$ photons.
If we fit this simulated spectrum with a single-temperature
{\small MEKAL} model, the resulting abundance is only about half
solar ($0.49\pm0.17$). The fit is statistically acceptable
(with $\chi^2=42.4/41$), partly due to the small counting statistics.
If the simulated spectrum contains $10^4$ photons instead,
the fit is no longer acceptable (with $\chi^2=213/91$)
and the best fit tends to give a much lower abundance
(about one-third solar).
A two-temperature component model can significantly improve the fit,
and the abundances are in general less than one solar,
although they are not strongly constrained.
Thus the abundance is likely underestimated
by fitting an X-ray spectrum of gas with broad temperature and density
distributions (see also \citealt{Strickland98}).

Given the low resolution and small signal-to-noise ratio of
X-ray spectra typically available for diffuse hot gas,
it is difficult to distinguish gas with a broad temperature distribution
from an isothermal plasma, especially within a narrow energy band.
In Fig.~\ref{F:spectrum} we plot an arbitrarily normalized
spectrum of a single-temperature hot plasma (0.8\,keV for Model A
and 0.4 keV for Model B,
where the EM distribution of the corresponding 1D model peaks).
The spectra of the 3D simulation and the isothermal plasma model
are similar in the 0.5--2.0\,keV band.
This approximation gives a potential shortcut to fit the X-ray
observations coarsely with only one or two gas components,
which are not necessarily able to reveal the actual physical and chemical state
of the ISM. With higher resolution spectra, line diagnostics
may provide additional useful information to reveal the gas properties.

However, the models still predict far lower X-ray luminosities
than observed. The observed diffuse X-ray luminosity 
in the 0.5--2~keV band of our Galactic bulge
and M\,31 bulge is about $10^{38}\rm~ergs~s^{-1}$
\citep{Shirey01,TOKM04,Muno04,Li07a}.
Using the reference mass and energy input rates,
the bolometric X-ray
luminosity predicated by the 1D wind model is no more than
$10^{36}\rm~ergs~s^{-1}$. The inhomogeneous
structures of the gas density and temperature in Model A
increase by no more than half an order of magnitude
the luminosity in the 0.5-2\,keV band (Table 2).
This under-prediction of the observed luminosity remains
a serious problem for the models.

\begin{figure}[htbp]
\begin{center}
\plotone{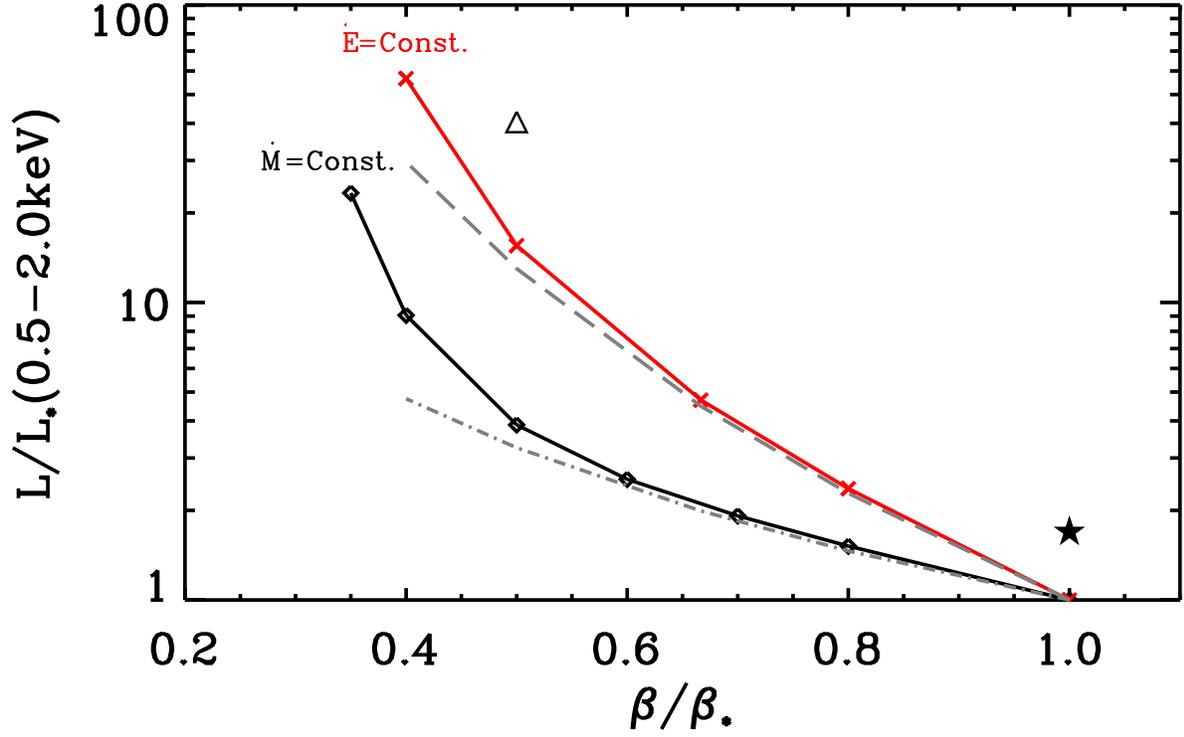}
\caption{\label{F:en_vs_em}
Dependence of the 0.5-2.0 keV luminosity on $\beta$.
The linked diamonds denote the results with a fixed $\dot{M}$,
while the crosses mark the results with a fixed $\dot{E}$.
The 3D simulations results are marked by the star for Model A
and the triangle for Model B.
Gray lines denote the corresponding scaling relationship (see \S4.3).
}
\end{center}
\end{figure}

The parameter with the strongest influence on the luminosity
of the bulge wind is the stellar mass loss rate.
As demonstrated in Model B which has a doubled mass input rate,
its mean gas density increases while the temperature decreases.
The X-ray emission
increases by a factor of $\sim 20$ in the 0.5-2.0\,keV band.
This enhancement is due to the increase of both density and emissivity.
To quantitatively understand this trend, let us first consider
a scaling relation of the bulge wind.
The specific heating $\beta = \dot{E}/\dot{M}$ determines
the mean central gas temperature of the wind solution.
In a steady flow, the velocity $u \propto \beta^{0.5}$
(e.g. \citealt{WC1983}), so long as gravitational potential
energy remains unimportant.
We have $\rho \propto \dot{M} \beta^{-0.5}$ from
the mass conservation equation,
and $T \propto \beta$ from the energy conservation equation.
For gas with a temperature between 0.5 and 1.0\,keV,
the X-ray emissivity in the 0.5--2.0\,keV band
is roughly inversely proportional to the temperature
[e.g., $\Lambda(T) \propto T^{-0.7}$ based on the {\small MEKAL} model].
The total X-ray emission is then $L\propto \rho^2 \Lambda(T)
\propto \dot{M}^{3.7} \dot{E}^{-1.7}$.
For winds with the same $\beta$ the velocity and temperature
profiles are the same, and density profiles differ only in their
normalization, which is proportional to $\dot{M}$.
We thus use $\beta/\beta_*$
to denote the separate change of either $\dot{M}$ or $\dot{E}$,
e.g, $L\propto \dot{M}^{3.7}$ is equivalent to
$L\propto (\beta/\beta_*)^{-3.7}$ for a fixed energy input,
and $L\propto \dot{E}^{-1.7}$ to $L\propto (\beta/\beta_*)^{-1.7}$
for a fixed mass input.
Fig.~\ref{F:en_vs_em} shows the luminosity in the 0.5--2 keV
as a function of $\beta$ based on a suite of 1D simulations.
In this plot the luminosities are normalized to that of a
1D reference model ($L_*\sim 10^{36}\rm ergs\,s^{-1}$;
corresponding to Model A).
The scaling relations of $L$ versus $\dot{M}$ and $\dot{E}$
closely match the simulations when $\beta/\beta_*$ is larger than 0.5.
In the adopted galactic bulge,
if the specific energy approaches one-third of the reference value,
the simulated luminosities increase sharply and no longer follow the
scaling relation, because the gravitational potential becomes
dynamically important for such a small $\beta$.
The corresponding results of Models A and B are also plotted in
Fig.~\ref{F:en_vs_em} to show the effects of gas inhomogeneity.
The ratio of the 0.5-2.0 keV luminosity of Model A
to that of the corresponding 1D model is about 2, 
and the ratio is about 3 for Model B.

Let $\beta$ be fixed at the reference value in Model A,
the velocity and temperature profiles of the wind should
remain the same and $L_X \propto \dot{M}^2$.
Hence, both $\dot{E}$ and $\dot{M}$
need to increase by an order of magnitude in order to
boost the X-ray luminosity to match that observed in the M\,31 bulge.
Although a bulge wind with a reduced $\beta$ can significantly increase
the diffuse X-ray emission as demonstrated in Model B,
it is unlikely to be the best solution.
In the presence of a reasonable dark matter halo with
properties like that of the Milky Way galaxy (e.g.,
$M_{halo} \simeq 10^{12}\msun$, $r_{virial} \simeq 250$\,kpc,
an NFW profile with a concentration parameter of 15),
$\sim 120\%$ of the available energy input in Model B
is required for the escape of all the stellar ejecta
to beyond the virial radius. Thus the bulge wind in Model B
is not energetic enough to escape the galaxy potential well
and the bulge wind material must accumulate within
the virial radius. This accumulated material may form
a considerable circum-galactic medium (CGM) that could
eventually quench the bulge wind \citep{Tang08}.

The gas outflow might be subsonic in the vicinity of galactic
bulges and other ellipticals with low $L_X/L_B$ ratios
under the interaction between the bulge wind and CGM.
The subsonic state is necessary to explain both the
large luminosity (compared to the wind model prediction)
and the extent of the diffuse X-ray emissions.
Even if a bulge wind has the energy to escape the halo
virial radius, the wind can be reversely shocked and stalled by the CGM.
When the reverse shock propagates inward within the sonic point,
the bulge wind turns into a globally subsonic outflow.
Such a subsonic state can be quasi stable for a long time
with proper treatment of the boundary the galactic flow \citep{Tang08}.
A proper 3-D simulation including the entire CGM is possible,
though computationally very expensive at present.

Other effects, which are ignored here for simplicity,
can further affect the X-ray luminosity of the galactic bulges.
A vertically collimated bulge wind can be a promising way to
significantly boost the X-ray luminosity, particularly for objects
such as M\,31 with a considerable mass in the galactic disk.
Motivated by the observed bipolar diffuse X-ray emission in the M\,31
bulge \citep{Li07a}, we have found in preliminary simulations that a
vertically collimated wind, confined by the surrounding gaseous disk,
will have significantly higher X-ray emission
(by up to an order of magnitude).
The major reason is that the density diverges much slower in the
collimated wind than in the spherical wind.
We plan to address this issue in a separate paper
(Tang \& Wang in preparation).

\section {Summary}\vspace{-\parskip}

In this work, we have explored the properties of the structured
hot gas created by sporadic SN explosions inside a galactic bulge
by conducting detailed 3D hydrodynamical simulations.
Our main results are as follows:
\begin{itemize}
\item A galactic wind may be generated in a galactic bulge
  with the standard empirical stellar mass loss and Type Ia SN rate.
  The gas properties fluctuate in time particularly
  in the central region lying within the sonic radius of the wind,
  where individual SN explosions strongly influence
  the density and temperature distributions.
  At larger radii, the spherically average profiles of the 3D simulations
  follow those of the 1D models. Therefore the 1D treatment of
  a galactic bulge wind flow is a reasonable approximation
  on large scales.

\item Sporadic SN explosions produce 3D filamentary and
  shell-like structures in the gas.
  These structures result in broad density and temperature
  distributions, compared to the 1D model. Furthermore,
  the relatively low temperature of the structures leads
  to an emission measure-weighted temperature that is
  significantly lower than the expected value inferred
  from the specific heating and has a relatively flat
  radial distribution throughout the bulge region,
  consistent with observations.

\item Iron ejected by SNe does not mix well with the surrounding gas
  within the bulge region and has a relatively high temperature
  and low density, so it contributes primarily to emission in the
  energy band $>$ 2\,keV. The diffuse soft X-ray emission
  comes from shells associated with SN blastwaves,
  which are hardly enriched by SN ejecta and have a metallicity
  close to the ambient gas that originates in stellar winds.
  This, together with the temperature broadening, helps to explain the
  apparent sub-solar abundance of the soft X-ray-emitting gas
  in galactic bulges/ellipticals.
 
\item
  Compared with the 1D spherical wind model,
  the structured hot gas in 3D simulations can boost
  the X-ray emission in an intermediate energy band
  (e.g., 0.5-2.0\,keV) by a factor of a few.
  This increase is more significant at the lower or
  higher energy bands due to the broad distributions
  of both temperature and density.
  In order to produce the luminosity and surface
  brightness distribution similar to the observed
  diffuse X-ray emission, the bulge outflow likely
  needs to be in a subsonic state and/or
  an angularly confined configuration.
  
\end{itemize}

\vspace{-\parskip}
\section*{Acknowledgments}\vspace{-\parskip}
The software used in this work was in part developed
by the DOE-supported ASC / Alliance Center for Astrophysical
Thermonuclear Flashes at the University of Chicago.
Simulations were performed at the Pittsburgh Supercomputing Center
supported by the NSF.
This project is supported by
NASA through grant SAO TM7-8005X and NNG07AH28G.



\renewcommand{\baselinestretch}{1.0}
\epsscale{0.65}

\epsscale{0.99}

\epsscale{0.9}

\epsscale{0.5}

\epsscale{1.0}

\epsscale{1.0}

\epsscale{1.0}

\epsscale{0.4}

\epsscale{0.450}

\epsscale{0.5}

\epsscale{0.65}

\epsscale{0.55}

\epsscale{1.0}

\epsscale{0.45}


\epsscale{1.0}

\epsscale{1.0} 

\epsscale{0.65}

\clearpage 

\epsscale{0.6}

\epsscale{1.0}
 

\clearpage

\appendix
\section{SNR Embedding Scheme}
\newcommand{\MVX}{M\hspace{-0.22em}V\hspace{-0.22em}X}
\newcommand{\MVY}{M\hspace{-0.22em}V\hspace{-0.22em}Y}
\newcommand{\MVZ}{M\hspace{-0.22em}V\hspace{-0.22em}Z}
\newcommand{\EK}{E\hspace{-0.15em}K}
\newcommand{\EI}{E\hspace{-0.14em}I}

After all quantities of zones within the affected region
are replaced by the dynamically generated SNR profiles,
we make additional adjustments to conserve the mass, momentum, and energy,
accounting for the geometrical effect due to discrete zones
and the possible net bulk motion of gas in that region.
Some symbols are introduced in the following adjustments for clarity: 
``M" denotes the total mass;
``\MVX'', ``\MVY'' and ``\MVZ'' are momentum in X, Y and Z directions, respectively; 
``\EK'' is the total kinetic energy;
``EI'' is the total thermal energy;
 the subscripts ``0'' and ``1'' denote the values before and after 
 the SNR seed is planted, 
 ``2'' stands for the adjusted values. The symbols followed by $(i,j,k)$ denote the
 values at each zone. In all the calculation a sub-zone method
 (e.g., subdividing a zone to determine
whether a zone is partially within the affected region)
is used in order to achieve high accuracy.
\newline
\noindent{\bf (a) Conservation of mass.}
It is accomplished by simply multiplying a factor, 
$f_{dens} = \rm M_2 /M_1 = ({M_0 + M_{ejecta} })/{M_1} $, to the density of
scaled SN profile:
\begin{equation}
\rho_{_2}(i,j,k) = f_{dens} \times  \rho_{_1}(i,j,k) \;.
\end{equation}
\begin{equation}
{\rm M}_0 =\sum_{i,j,k} \rho_{_0}(i,j,k)\Delta V_{i,j,k}
\end{equation} 
\begin{equation}
{\rm M}_1 =\sum_{i,j,k} \rho_{_1}(i,j,k)\Delta V_{i,j,k}
\end{equation}
\vspace*{-0.5em}
$M_{ejecta}$ denotes the SN Ia ejecta and is 1.4 $M_\odot$.
The summation ($\Sigma_{i,j,k}$) is over the SNR affected regions.
Note that the change of mass will also affect the conservations
of other values such as momentum and energy so this step must
be done first. In general, $f_{dens}$ is very close to one,
which is guaranteed by the dynamically generated SNR profile.
\newline
\noindent{\bf (b) Conservation of momentum.}
There is no net momentum in each direction for the SNR seeds.
However, the region to embed the SNR may contain non-zero momentum,
especially when an SNR actually explodes in an outflowing wind region.
Taking the momentum in the X direction as an example: 
the initial net momentum in X direction is
\begin{equation}
{\rm \MVX_0} = \sum_{i,j,k}\rho_{_0}(i,j,k) v_{x,0}(i,j,k) \Delta V_{i,j,k}\;, 
\end{equation}
In order to conserver the momentum, we add a bulk motion 
\begin{equation}
\overline{v}_{x,1} = \frac{\rm \MVX_0}{\rm M_2} 
\end{equation}
to the embedded SNR velocity profile
\begin{equation}
v_{x,1}(i,j,k) = v_{x,sn}(i,j,k) + \overline{v}_{x,1} \;.
\end{equation}
\begin{equation}
  {\rm \MVX_1}=  \sum_{i,j,k} \rho_{2}(i,j,k) v_{x,1}(i,j,k)\Delta V_{i,j,k} \;,
\end{equation}
Here $v_{x,sn}$ is the velocity profile of the SNR template.
Let $f_{vx} = \rm \MVX_0 / \MVX_1$, thus 
\begin{equation}
v_{x,2}(i,j,k) = f_{vx}\times v_{x,1}(i,j,k) \;.
\end{equation}
The same procedure is also applied to Y and Z direction.
\newline
\noindent{\bf (c) Conservation of energy.}
The kinetic energy of the region is
\begin{equation}
\EK_0 = \frac{1}{2}\sum_{i,j,k} \rho_{0}(i,j,k) \bigg( v_{x,0}^2(i,j,k) +
v_{y,0}^2(i,j,k) + v_{z,0}^2(i,j,k) \bigg) \Delta V_{i,j,k}
\end{equation}
After adjusting the mass and momentum, the updated kinetic energy is:
\begin{equation}
\EK_2 = \frac{1}{2}\sum_{i,j,k} \rho_{2}(i,j,k) \bigg( v_{x,2}^2(i,j,k) +
v_{y,2}^2(i,j,k) + v_{z,2}^2(i,j,k) \bigg)\Delta V_{i,j,k} \;.
\end{equation}
The energy conservation is achieved by adjusting 
the thermal energy accordingly:
\begin{equation}
p_2(i,j,k) = f_{_{\EI}} p_1(i,j,k),
\end{equation}
where
\begin{equation}
f_{_{\EI}} = \frac{\EI_0 + EK_0 + E_{sn} - \EK_2}{\EI_1},\ \ 
\EI_0 = \sum_{i,j,k} \frac{p_0(i,j,k)}{\gamma-1} \Delta V_{i,j,k},\ \ 
\EI_1 = \sum_{i,j,k} \frac{p_1(i,j,k)}{\gamma-1} \Delta V_{i,j,k}.
\end{equation}
Note the original random motion of mass inside the affected 
region is implicitly converted to thermal energy.
The clumpy environment is also smoothed out by this scheme,
and the embedded SNR is spherical symmetry related to the SNR center.


\begin{thebibliography}{} 
\setlength{\itemsep}{-0.25truecm}

\bibitem[Athey \etal(2002)]{Athey02}
Athey A., Bregman J., Bregman J., Temi P., Sauvage M., 2002, ApJ, 571, 272

\bibitem[Borkowski \etal(2001)]{Borkowski01}
Borkowski K. J., Lyerly W. J., Reynolds S. P. 2001, \apj, 548, 820

\bibitem[Blum(1995)]{Blum1995}
Blum, R. D. 1995, \apj, 444, L89

\bibitem[Canto et~al.(2000)]{Canto00}
Canto J., Raga A. C., Rodriguez L. F., 2000, ApJ, 536, 896

\bibitem[Cappellaro \etal(1999)]{CappEva1999}
Cappellaro E., Evans R., Turatto M. 1999, \aap, 351, 459

\bibitem[Ciotti \etal(1991)]{Ciot91}
Ciotti L., Pellegrini S., Renzini A., \& D'Erocole A. 1991, ApJ, 376, 380

\bibitem[Cox (2000)]{Cox00}
Cox A. N., 2000, Allen's Astrophysical Quantities, fourth edition.

\bibitem[David et~al.(2006)]{David06}
David L. P., Jones C., Forman W., Vargas I. M., Nulsen P., 2006, ApJ, 653, 207

\bibitem[de Avillez \& Breitschwerdt(2005)]{Avillez05}
de Avillez M. A., \& Breitschwerdt D., 2005, A\&A, 436, 585

\bibitem[Eckart \etal(1993)]{Eckart1993}
Eckart A., Genzel R., Hofmann R., Sams B. J., Tacconi-Garman L. E.,
1993, ApJ, 407, L77

\bibitem[Fryxell \etal(2000)]{Fryxell00}
Fryxell B., et al., 2000, APJS, 131, 273

\bibitem[Hernquist(1990)]{Hernquist1990}
Hernquist L. 1990, \apj, 356, 359
	
\bibitem[Hnatyk \& Petruk (1999)]{Hnatyk99}
Hnatyk B., Petruk O. 1999, A\&A, 344, 295

\bibitem[Humphrey \& Buote(2006)]{Humphrey06}
Humphrey P. J., \& Buote D. A., 2006, ApJ, 639, 136

\bibitem[Irwin \etal(2002)]{Irwin02}
Irwin, J. A., Sarazin, C. L., Bregman. J. N. 2002, ApJ, 570, 152

\bibitem[Joung \& Mac Low(2006)]{Joung06}
Joung M. K., \& Mac Low M.-M., 2006, ApJ, 653, 1266

\bibitem[Joung et~al.(2008)]{Joung08}
Joung M. K., Mac Low M.-M., Bryan G. L., 2008, astroph/arXiv0811.3747

\bibitem[Kent(1992)]{Kent1992}
Kent S. M. 1992, \apj, 387, 181

\bibitem[Lepine \& Leroy(2000)]{LepineL00}
Lepine J. R. D. \& Leroy P. 2000, \mnras, 313, 263

\bibitem[Li \& Wang (2007)]{Li07a}
Li Z.,  Wang Q. D., 2007, ApJ, 668, L39

\bibitem[Li et~al.(2007)]{Li07b}
Li Z., Wang Q. D., Hameed S. 2007, MNRAS, 376, 960

\bibitem[Liedahl et~al.(1995)]{Liedahl95}
Liedahl D. A., Osterheld A. L., \& Goldstein W. H., 1995, ApJ, 438, L115

\bibitem[Lowenstein \& Mathews(1987)]{LM1987}
Lowenstein M., \& Mathews W. 1984, \apj, 319, 614

\bibitem[L$\rm\ddot{o}$ner (1987)]{Lohner87}
L$\rm\ddot{o}$hner R. 1987, Comp. Meth. App. Mech. Eng., 61, 323 

\bibitem[Mac Low et~al.(2005)]{MacLow05}
Mac Low M.-M., Balsara D. S., Kim J., de Avillez M. A., 2005, ApJ, 626, 864

\bibitem[Maraston(2005)]{Mar05}
Maraston C. 2005, \mnras, 362, 799

\bibitem[Mathews \& Baker(1971)]{MB1971}
Mathews W. G. \& Baker J. C. 1971, \apj, 170, 241

\bibitem[Mewe et~al (1985)]{Mewe85}
Mewe R., Gronenschild E. H. B. M., \& van den Oord G. H. J. 1985, A\&AS, 62, 197

\bibitem[Muno \etal(2004)]{Muno04}
Muno M. P., et al., 2004, ApJ, 613, 326

\bibitem[Nomoto \etal(1984)]{Nomoto1984}
Nomoto L., Thielemann F., \& Yokoi L. 1984, \apj, 286, 644

\bibitem[O'Sullivan \etal(2003)]{OEPT03}
O'Sullivan E., Ponman T. J., Collins R. S. 2003, MNRAS, 340, 1375

\bibitem[Sage et~al. (2007)]{Sage07}
Sage L. J., Welch G. A., Young L. M. 2007, ApJ, 657, 232

\bibitem[Sarazin et~al.(2001)]{Sarazin01}
Sarazin C. L., Irwin J. A., \& Bregman J. N. 2001, \apj, 556, 533.

\bibitem[Sato \& Tawara (1999)]{Sato99}
Sato S., \& Tawara Y. 1999, \apj, 514, 765

\bibitem[Sedov (1959)]{Sedov59}
Sedov L. I 1959, Similarity and Dimensional Methods in Mechanics,
translation from 4th Russian edition, Academic press New York and London

\bibitem[Shirey \etal(2001)]{Shirey01}
Shirey R., et al., 2001, A\&A, 365, L195

\bibitem[Shu(1992)]{Shu1992}
Shu F. H. 1992, The physics of astrophysics, Volume II, Gas Dynamics 
(University Science Books)

\bibitem[Strickland \& Stevens(1998)]{Strickland98}
Strickland D. K., \& Stevens I. R., 1998, MNRAS, 297, 747

\bibitem[Sutherland \& Dopita(1993)]{Sutherland93}
Sutherland R. S., \& Dopita M. A., 1993, APJS, 88, 253

\bibitem[Tang \& Wang(2005)]{Tang05}
Tang S., \& Wang Q. D. 2005, ApJ, 628, 205

\bibitem[Tang \etal (2008)]{Tang08}
Tang S., Wang Q. D., Lu Y., Mo H. J., 2009, MNRAS, 392, 77

\bibitem[Tang \& Wang  (2009)]{Tang09} 
Tang S., \& Wang Q. D.,  2009, astroph/arXiv0902.0403

\bibitem[Takahashi \etal(2004)]{TOKM04}
Takahashi H., Okada Y., Kokubun M., Makishima K., 2004, ApJ, 615, 242

\bibitem[Wang (2007)]{Wang07}
Wang Q.~D. 2007, EAS, 24, 59

\bibitem[White \& Chevalier(1983)]{WC1983}
White R. E., \& Chevalier R. A. 1983, \apj, 275, 69

\bibitem[Wolfire \etal(1995)]{Wolfire1995}
Wolfire M. G., McKee C. F., Hollenbach D., Tielens A. G. G. M.,
1995, ApJ, 453, 673

\bibitem[Zhao(1994)]{Zhao1994}
Zhao, H. S. 1994, Ph.D. thesis, Columbia Univ.

\end{thebibliography}
\end{document}